\documentclass[11pt]{article}
\usepackage{jcappub}

\makeatletter

\usepackage{graphicx}

\newcommand{\sun}{\ensuremath{\odot}}

\def\ln{{\rm ln}}

\usepackage{hyperref}
\usepackage{esint}

\title{UHECR Acceleration in Dark Matter Filaments of Cosmological Structure
Formation }

\author[a]{M. A. Malkov,}
\author[b]{R.Z. Sagdeev} 
\author[a,c]{and P.H. Diamond}

\affiliation[a]{CASS and Department of Physics, University of California,
San Diego, La Jolla, CA 92093-0424}

\affiliation[b]{University of Maryland, College Park, MD 20742-3280}
\emailAdd{mmalkov@ucsd.edu}

\affiliation[c]{WCI Center for Fusion Theory, National Fusion Research Institute,
Gwahangno 113, Yuseong-gu, Daejeon 305-333, Republic of Korea}

\abstract{

A mechanism for proton acceleration to $\sim10^{21}eV$ is suggested.
It may operate in accretion flows onto thin dark matter filaments
of cosmic structure formation. The flow compresses the ambient magnetic
field to strongly increase and align it with the filament. Particles
begin the acceleration by the $\mathbf{E}\times\mathbf{B}$ drift with
the accretion flow. The energy gain in the drift regime is limited
by the conservation of the adiabatic invariant $p_{\perp}^{2}/B$$\left(r\right)$.
Upon approaching the filament, the drift turns into the gyro-motion
around the filament so that the particle moves parallel to the azimuthal
electric field. In this 'betatron' regime the acceleration speeds
up to rapidly reach the electrodynamic limit $cp_{max}=eBR$ for an
accelerator with magnetic field $B$ and the orbit radius $R$ (Larmor
radius). The
periodic orbit becomes unstable and the particle slings out of the
filament to the region of a weak (uncompressed) magnetic field, which
terminates the acceleration. 

To escape the filament, accelerated particles must have gyro-radii
comparable with the filament radius. Therefore, the mechanism requires
pre-acceleration that is likely to occur in structure formation shocks
upstream or nearby the filament accretion flow. Previous studies identify
such shocks as efficient proton accelerators to a firm upper limit
$\sim10^{19.5}eV$ placed by the catastrophic photo-pion losses. The
present mechanism combines explosive energy gain in its final (betatron)
phase with prompt particle release from the region of strong magnetic
field. It is this combination that allows protons to overcome both
the photo-pion and the synchrotron-Compton losses and therefore attain
energy $\sim10^{21}eV$. A customary requirement on accelerator power
to reach a given $E_{{\rm max}}$, which is placed by the accelerator
energy dissipation $\propto E_{{\rm max}}^{2}/Z_{0}$ due to the finite
vacuum impedance $Z_{0}$, is circumvented by the cyclic operation of the accelerator.
}

\arxivnumber{0000.0000}
\begin{document}
\maketitle

\section{Introduction}

A relatively strong, few $\mu{\rm G}$ galactic magnetic field precludes
tracing Galactic cosmic rays (CR) back to their origins due to the
magnetic orbit scrambling. Therefore, the long suspected link between
CRs and supernova remnants (SNR) for example, even when convincingly
established, will be indirect. The extra-galactic ultra-high energy
CRs (UHECR), on the contrary, are open to what is called {}``astronomy
with charged particles'' (e.g., \citep{SiglMiniatiEnsslin03,Waxman09}).
Indeed, a $10^{20}eV$ proton has a 100 Mpc gyro-radius in a nano-Gauss
intergalactic magnetic field. These particles propagate almost rectilinearly
and may thus point back to the locations of the most sophisticated
nature's accelerators. From the theory side, however, there is as
yet no astrophysical object which could be considered as an unquestionable
candidate for acceleration and subsequent release of particles with
such extreme energies.

An important recent progress in the emerging field of CR astronomy
was made by the Auger team who were able to find the UHECR correlation
with the large scale structure, as it traced by the AGN \citep{Auger07Sci}.
Some of the earlier studies also point at the UHECR clustering, e.g.,
\citep{TakedaClustering99}, while other analyses incline towards
a more isotropic distribution (see, e.g. \citep{KrishnaBier10,KoteraOlinto11} for 
a review of the recent results and \citep{PAO_anis10} for the
updated Auger observations, downplaying to some extent the initial
anisotropy finding). Notwithstanding the recent advances, the origin
of the mysterious UHECRs remains unknown. 

A necessary step for identifying the UHECR sources is to test the
putative extragalactic accelerators for the capability to accelerate
protons beyond $10^{20}$eV. It seems natural to attempt at applying
well advanced galactic CR acceleration mechanisms to such accelerators.
The popular diffusive shock acceleration (DSA) is widely accepted
as the most promising mechanism to accelerate galactic CRs. Possible
sites of its application to the UHECR acceleration are the cosmic
structure formation shocks \citep{KangCluster96}, gamma ray bursts
\citep{MilgromUsov95,Vietri95,Wax95} and the AGN environments \citep{Farrar98,Bland00}.
They have been examined in a number of publications, e.g., \citep{NormanAchterb95,Bland00,AharPRD02,JonesUHECR04,Torres04}.
Remarkably, the DSA falls short by one order of magnitude to produce
particles in the structure formation shocks with the highest energy
observed, i.e. a few $10^{20}$ eV \citep{NormanAchterb95,JonesUHECR04},
as the other scenarios encounter serious problems as well. At the
same time, the DSA seems to be quite capable of accelerating particles
to $\sim10^{19.5}$eV. However, while promising attempts to incorporate
the CR-acceleration into simulations of the large-scale structure
formation have been made \citep{Miniati01,KangUHECR04,KangJones05,Ensslin07,Pfrommer07},
each of these two processes is scarcely feasible computationally,
if all essential phenomena are included.

The purpose of this paper is to demonstrate that protons accelerated
in the structure formation shocks to $\sim10^{19.5}$eV can be boosted
to $10^{21}$eV inside the same accretion flow. The suggested mechanism
accelerates particles much faster than the DSA, thus sustaining against
losses. It operates in plasmas accreting on to the gravitating dark
matter (DM) filaments. Filaments, along with pancakes and knots are
important elements of the cosmic structure formation which was established
in a number of simulations (e.g., \citep{Shandarin96,RyuKangTW03,KangUHECR04,KangJones05,CenOstriker06,2010NewA...15..695V})
and seems to be supported observationally \citep{ColdFilaments06,FilamentObs08}.
The physical reason for the high acceleration rate of this mechanism
can be readily understood by comparing it with the scatter-free shock
surfing or shock drift acceleration mechanisms operating in perpendicular
shocks \citep{SagdeevUppsala60,Schatzman63,LeeSS96,Zank96}. Note,
however, that the present mechanism does not require a shock in the
flow, so a gradual flow compression suffices. When a rapidly gyrating
high energy particle is slowly convected with the flow speed $u\ll c$
through a shock, the shock-particle interaction is adiabatic. The
reason for that is a small $\sim u/c\ll1$ momentum gain after each
shock crossing and re-crossing. After $c/u\gg1$ such cycles a particle
of momentum $p_{\perp}$ is left downstream with the net energy gain
$\sim1$, or more precisely, to conserve the adiabatic invariant $p_{\perp}^{2}/B$,
where $B$ is the magnetic field (e.g., \citep{Toptygin80,Drury83}).
The scatter-free mechanism works $\sim c/u\gg1$ times faster than
the DSA, since the upstream and downstream residence time is only
of the order of gyro-period $\omega_{c}^{-1}$, as opposed to the
$c/u\omega_{c}$ idling time of the DSA. However, the duration of
the scatter-free acceleration is too short to reach high energies.
It should be noted that under certain circumstances its operation
can be prolonged. For example, the cross shock potential can retain
\emph{nonrelativistic }particles at the shock against convection downstream.

In the mechanism proposed in this paper the role of retaining force
plays the centrifugal potential that efficiently keeps particles against
convection into the filament when the filament radius is sufficiently
small. Initially, however, a particle is also convected by the flow
towards the filament. As long as the particle drifts towards the center
with the flow adiabatically, the magnetic moment $p_{\perp}^{2}/B$
is conserved, and the particle gains energy according to this relation.
At this phase of acceleration, the energy gain is relatively slow
since the gyro-averaged work done by the azimuthal (motion) electric
field approximately cancels out. However, if the particle has a gyro-radius
$r_{g}\gtrsim R_{{\rm f}}$ (filament radius), the particle adiabatic
invariant strongly deviates from $p_{\perp}^{2}/B$ before the particle
sinks into the filament. Particle motion changes from the slow drift
\emph{towards }the filament to a nearly circular motion \emph{around
}it. The effect of work cancellation completely disappears and the
particle enters into a betatron acceleration regime with a very fast
(explosive) energy gain. Only after the orbit radius exceeds a critical
value, the particle slings out of the filament vicinity. The acceleration
rate drops since the motion electric field far away from the filament
is weak and the acceleration virtually terminates.

The remainder of the paper is organized in two parts. The first, preparatory
part deals with the configuration of accretion flow where the acceleration
mechanism takes place, Sec.\ref{sec:Accretion-flow} and Appendix
\ref{sec:ApFlowAndMF}. Given the inability of the DSA to accelerate
protons beyond $10^{20}$eV, we focus on the principal possibility
of acceleration to such extreme energies. Therefore, in the first
part we limit our consideration to the most favorable for the acceleration,
simple flow. The second part of the paper (Sec.\ref{sec:Acceleration})
deals with a detailed analytic description of the proposed acceleration
mechanism. Ample numerical illustrations are also provided there.
We estimate energy losses in Sec.\ref{sec:Losses} and discuss other
limitations of the mechanism in Sec\ref{sec:Conclusions}.

\section{Accretion flow onto filaments and nodes in large scale structures\label{sec:Accretion-flow}}

Before describing the particle acceleration mechanism in the cosmological
structures, it is necessary to understand the geometry of magnetic
and motion electric fields generated by the accreting plasma. A useful
guidance is provided by numerical simulations performed within the
$\Lambda{\rm {CDM}}$ model \citet{CenOstriker06,Dolag05,RyuKangTW03,MilleniumII09,Miniati00}.
In such simulations, the gravitationally interacting dark matter (DM)
particles aggregate to form a structure which then gravitationally
drives conducting gas with the frozen in magnetic field. We will neglect
the self-gravitation of the gas as it comprises only a relatively
small fraction of the total mass. The emerging structures and thus
the induced plasma flows are complicated in detail but morphologically
generic. The matter accrets onto sheets, filaments and nodes and is
thus organized in a {}``cosmic web'' of massive nodes connected
by relatively dense filaments along which the matter flows towards nearby nodes.
The rest of the space can be considered as low density, low magnetic
field {}``voids''. 

Returning to the particle acceleration in such structures, we focus
on a single filament with two nodes at its ends (a 'dumbbell'), Fig.\ref{fig:Dumbbell}.
The dumbbell structure is supported by the DM accretion which is briefly
discussed in \ref{sub:Dark-matter-gravitational}. To set up the acceleration
scheme we note that the strong flow compression near the knots creates
magnetic mirrors that confine energetic particles in the field-filament
direction. A rarefied plasma accrets onto the filament from the
surrounding void and stream then towards the nodes, while partially
the plasma accrets onto the nodes directly from the void. The reaction
from the magnetic field is negligible, so that we can split our task
of setting out the magnetized flow configuration in the following
two parts. First, we evaluate velocity and density distribution in
the flow driven by the DM gravity, \ref{sub:Plasma-flow-towards}.
Second, we determine the magnetic field as being passively advected
by the accreting gas from the ambient medium, \ref{sub:Magnetic-field-around}.
According to the above consideration of the distribution of the magnetic
and electric fields around a filament, it is not unreasonable to assume
that, at least in the case of a favorable magnetic field orientation
far away from the filament, the field is well aligned with the filament
in the active acceleration zone. This assumption will be further discussed
in Sec.\ref{sec:Conclusions}.

\section{Particle acceleration around a filament\label{sec:Acceleration}}

Based on the flow pattern discussed above, we consider a DM filament
of radius $R_{{\rm f}}$ that accrets intergalactic gas in radial
direction. We assume that the ambient magnetic field inside of an
effective accretion volume (Bondi radius $R_{{\rm B}}$) is aligned
with the filament. We specify then $\mathbf{B}$ as $\mathbf{B}=\left(0,0,-B\right)$
with $B\left(r\right)$ depending only on $r=\sqrt{x^{2}+y^{2}}$,
the distance to the filament axis ($z$-axis), while particle motion
in $z$-direction is constrained by magnetic mirrors near the filament
end nodes. We do not include magnetic mirrors explicitly, but merely
assume that the dynamics of accelerated particles is nearly perpendicular
to $\mathbf{B}$, i.e. $p_{\parallel}\ll p_{\perp}\approx p$. This
assumption will be discussed in Sec.\ref{sec:Conclusions}. The equations
of motion in the polar coordinates $\left(r,\vartheta\right)$ on
the $\left(x,y\right)$ plane read

\begin{eqnarray}
\dot{p}_{r} & = & -\frac{p_{\vartheta}}{p}\left(eB-\frac{c}{r}p_{\vartheta}\right)\label{eq:pr}\\
\dot{p}_{\vartheta} & = & \frac{p_{r}}{p}\left(eB-\frac{c}{r}p_{\vartheta}\right)+eE_{\vartheta}\label{eq:pt}\\
\dot{r} & = & c\frac{p_{r}}{p}\label{eq:r}\\
\dot{\vartheta} & = & \frac{c}{r}\frac{p_{\vartheta}}{p}\label{eq:theta}\end{eqnarray}
where $p_{r}$ and $p_{\vartheta}$ are the radial and azimuthal components
of the particle momentum, $p=\sqrt{p_{r}^{2}+p_{\vartheta}^{2}}\gg m_{{\rm p}}c$
(with $m_{{\rm p}}$ being the proton rest mass), $r$ is the particle
radial coordinate, $\vartheta$ is the azimuthal angle, and $E_{\vartheta}$
is the azimuthal electric field. It is convenient to scale $B$ to
its value at infinity, $B_{\infty}={\rm const}$, both the radial
coordinate $r$ and the particle gyro-radius $r_{{\rm g}}\left(p\right)=pc/eB_{\infty}$
to $R_{{\rm B}}$ (which is the Bondi radius, eq.{[}\ref{eq:BondiRad}{]}),
and time, to $R_{{\rm B}}/c$. Thus, the particle momentum $p$ is
now measured in the units of $eB_{\infty}R_{{\rm B}}/c$. Since $E_{\vartheta}=-u_{r}\left(r\right)B\left(r\right)/c$,
where $u_{r}<0$ is the radial flow velocity, and since $ru_{r}B={\rm const}$,
the motion electric field $E_{\vartheta}\propto1/r$. It is thus suggestive
to introduce the following parameter 

\begin{equation}
v=-\frac{ru_{r}B}{R_{{\rm B}}cB_{\infty}}>0\label{eq:vpar}\end{equation}
that controls both the drift of energetic particles towards the filament
and their acceleration. Using these dimensionless variables (without
relabeling), eqs.(\ref{eq:pr}-\ref{eq:theta}) rewrite 

\begin{eqnarray}
\dot{p}_{r} & = & -\frac{p_{\vartheta}}{p}\left(B-\frac{1}{r}p_{\vartheta}\right)\label{eq:prn}\\
\dot{p}_{\vartheta} & = & \frac{p_{r}}{p}\left(B-\frac{1}{r}p_{\vartheta}\right)+\frac{v}{r}\label{eq:ptn}\\
\dot{r} & = & \frac{p_{r}}{p}\label{eq:rn}\\
\dot{\vartheta} & = & \frac{1}{r}\frac{p_{\vartheta}}{p}\label{eq:thetadot}\end{eqnarray}
Note that due to the azimuthal symmetry, the angular variable $\vartheta$
is ignorable and only the first three equations need to be solved
as a system. The angular variable $\vartheta$ is, however, useful
in that it traces the particle energy. Indeed, by virtue of eqs.(\ref{eq:prn}-\ref{eq:ptn})

\begin{equation}
\dot{p}=\frac{vp_{\vartheta}}{rp},\label{eq:pdot-1}\end{equation}
so that from eq.(\ref{eq:thetadot}) we obtain

\begin{equation}
p-v\vartheta=\mathrm{const.}\label{eq:pthetarelation}\end{equation}
Furthermore, the azimuthal component of the particle canonical momentum 

\begin{equation}
\mathcal{P}\left(t\right)=\Psi-rp_{\vartheta}\label{eq:canmomdef}\end{equation}
decreases linearly with time:

\begin{equation}
{\cal \mathcal{P}}+vt=\mathrm{const}\label{eq:canmomint}\end{equation}
where $\Psi\left(r\right)$ is defined as follows

\begin{equation}
\Psi\equiv\int_{0}^{r}rBdr\label{eq:Psi}\end{equation}

Using eq.(\ref{eq:canmomdef}), it is convenient to reduce the dynamical
system given by eqs.(\ref{eq:prn}-\ref{eq:rn}) to the following
1D Hamiltonian system

\begin{eqnarray}
\dot{p}_{r} & = & -\frac{\partial p}{\partial r}\label{eq:prham}\\
\dot{r} & = & \frac{\partial p}{\partial p_{r}}\label{eq:rham}\end{eqnarray}
where the particle momentum $p$ assumes the role of the Hamiltonian 

\begin{equation}
p\left(p_{r},r,t\right)=\sqrt{p_{r}^{2}+\left[\frac{\Psi\left(r\right)-\mathcal{P}\left(t\right)}{r}\right]^{2}}\label{eq:Hamilt}\end{equation}
Note that the time $t$ enters the Hamiltonian through $\mathcal{P}=\mathcal{P}_{0}-vt$.
If $v=0$, the relation $p=\mathrm{const}$ provides a complete solution
of the problem, at least in the form $t=t\left(r\right)$, seeing
that $p_{r}=p\dot{r}$.  We assume that $v\ll1$, i.e. the plasma
gravitational infall is slow compared to the speed of light. Therefore,
we may consider both the change in the particle {}``total energy''
$p^{2}$ and the deformation of the {}``potential energy'' of the
Hamiltonian in eq.(\ref{eq:Hamilt}) $U\left(r,t\right)\equiv\left(\Psi-\mathcal{P}\right)^{2}/r^{2}$
being adiabatic. 

\begin{figure}[h]
\includegraphics[bb=0bp 230bp 612bp 792bp,scale=0.7]{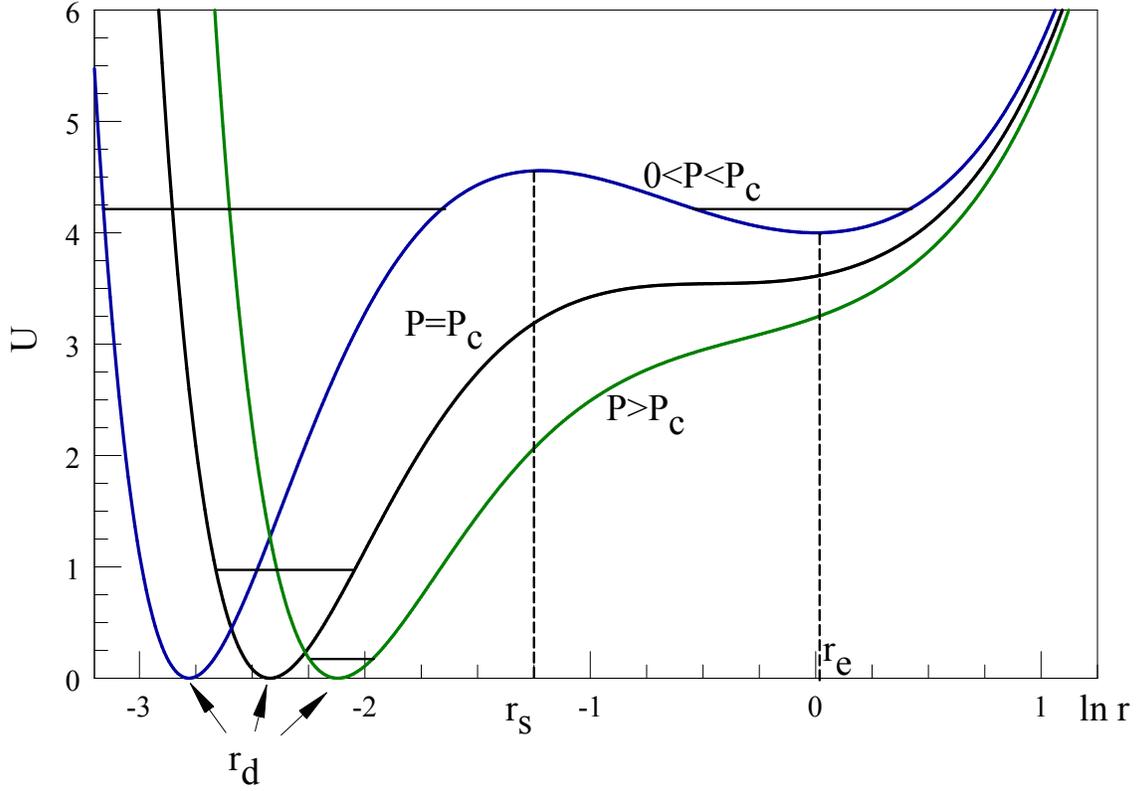}

\caption{Potential energy of a particle as a function of $r$, as it evolves
in time when $\mathcal{P}\left(t\right)$ varies from $\mathcal{P}\left(t\right)>\mathcal{P}_{c}$
to $\mathcal{P}\left(t\right)<\mathcal{P}_{c}$. Note that the radial
direction is given in a logarithmic scale, for clarity. In reality,
the potential well at smaller $r$ is much narrower than that at larger
$r$. The motion in the broad potential well corresponds to very large
orbit radius, which is equivalent to particle escape from the accelerator.\label{fig:Potential-energy-of}}

\end{figure}

\begin{figure}[h]
\includegraphics[bb=0bp 300bp 600bp 712bp,scale=0.7]{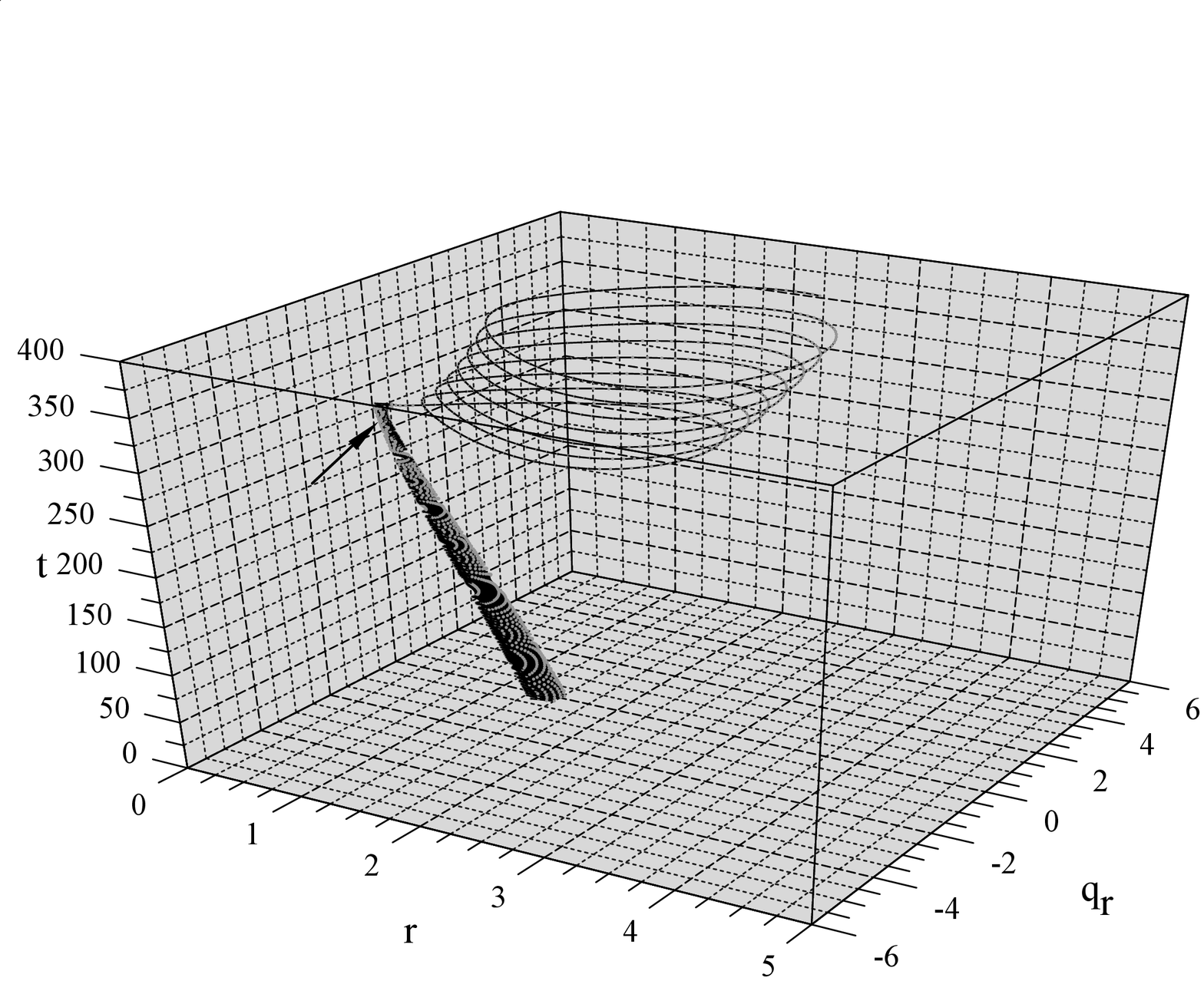}

\caption{3D particle trajectory with time pointing upward and with $\left(r,q_{r}\right)$
horizontal phase plane coordinates, where $q_{r}=rp_{r}$. The particle
starts its motion by drifting towards the filament from a distance
$r=1.5$, then it circulates around the filament (small $r$, also
shown with arrow) and, finally, it crosses the separatrix (see Fig.\ref{fig:Phase plane})
and becomes detrapped from the filament. The particle remain bounded
only by the ambient magnetic field (upper part of the trajectory)
\label{fig:3Dtrajectory}}

\end{figure}

\begin{figure}[h]
\includegraphics[bb=0bp 230bp 612bp 792bp,scale=0.7]{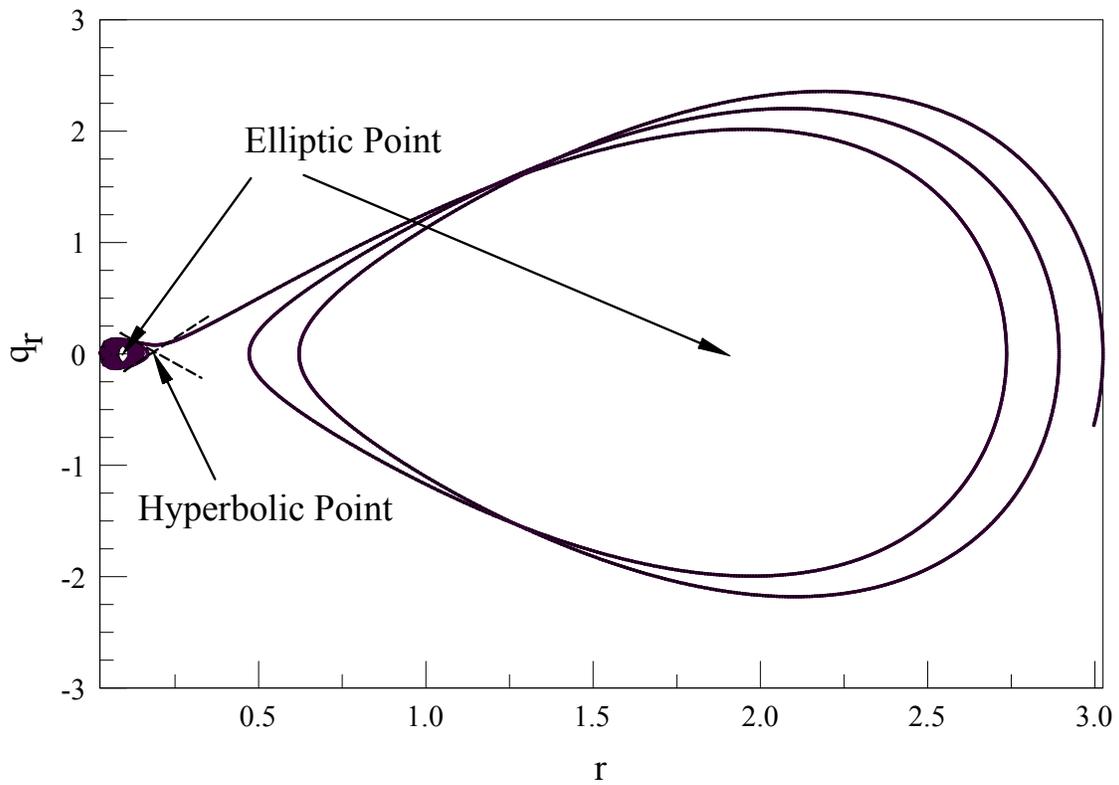}

\caption{Two parts of particle trajectory (shown in Fig.\ref{fig:3Dtrajectory})
after approaching the filament. The first part is to the left from
a hyperbolic point (betatron regime). The second part is when particle
crosses the separatrix and transitions to a weakly bounded state to
the right from the hyperbolic fixed point. Short pieces of the separatrix
passing through the hyperbolic point are shown with the dashed lines
\label{fig:Phase plane}}

\end{figure}

\begin{figure}[h]
\includegraphics[bb=0bp 130bp 612bp 792bp,scale=0.7]{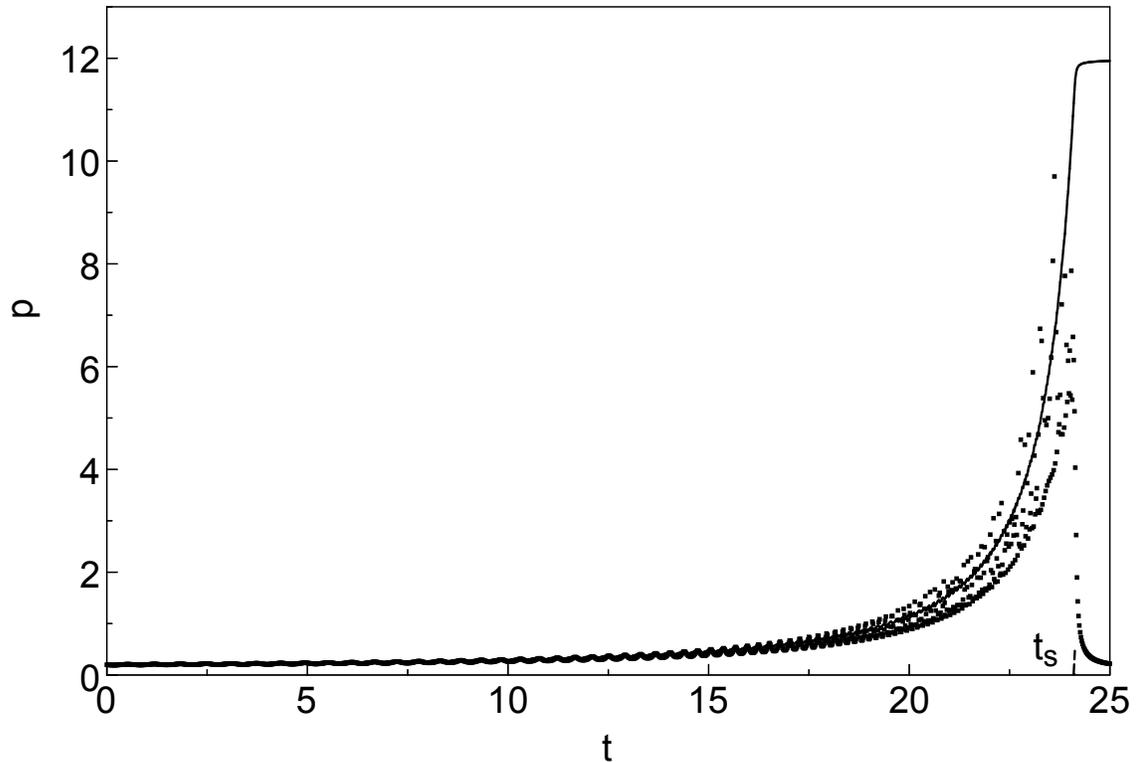}

\caption{Acceleration of a particle with the initial momentum $p_{0}=0.2$
and coordinate $r_{0}=1$ ($B_{0}=2$), $v=0.1$ (this case corresponds
to the rightmost point on the plot shown in Fig.\ref{fig:Particle-maximum-momentum}).
Solid line: numerical integration of the particle trajectory, shown
as $p\left(t\right)$. Dots: $p\left(t\right)$ obtained from the
condition $p^{2}/B\left(r\right)={\rm const}$ with 
$r\left(t\right)$ taken from the actual particle trajectory.
Significant spreading of points illustrates the inaccuracy of the above
representation of the adiabatic invariant near the separatrix. \label{fig:PoftWithAdiab} }

\end{figure}

Given the above considerations, the particle dynamics is easily understood
in terms of the critical points (i.e., points where $\partial U/\partial r=p_{r}=0)$
of the l.h.s. of eq.(\ref{eq:prham}-\ref{eq:rham}). One such point
is where $\Psi\left(r\right)=\mathcal{P}$. It definitely exists for
particles that move not too close to $r=0$ with $p$ being not too
large. Specific conditions will be given below. Evidently, this particular
critical point is the coordinate of the particle guiding center $r_{{\rm d}}$,
given by the following relation

\begin{equation}
\Psi\left(r_{{\rm d}}\right)=\mathcal{P}_{0}-vt.\label{eq:rdrift}\end{equation}
Apart from drifting with the velocity $\dot{r}_{{\rm d}}=-v/r_{{\rm d}}B\left(r_{{\rm d}}\right)$,
particles oscillate in the potential well near its minimum, Fig.\ref{fig:Potential-energy-of}.
The following adiabatic invariant, which can be calculated using general
principles of Hamiltonian dynamics, eqs.(\ref{eq:prham}-\ref{eq:Hamilt})

\begin{equation}
I=\oint p_{r}dr=\oint\sqrt{p^{2}-U\left(r,t\right)}dr\simeq\pi p^{2}/B\left(r_{{\rm d}}\right),\label{eq:adinv}\end{equation}
is approximately conserved. The last relation in eq.(\ref{eq:adinv})
is valid in the guiding center approximation, discussed above. Since
$\Psi\ge0$ for all $r$, the solution of eq.(\ref{eq:rdrift}) for
$r_{{\rm d}}$ ceases to exist for $t>\mathcal{P}_{0}/v$. Formally
this means that the particle guiding center reaches the origin, but
the particle itself does not. Clearly, the guiding center approximation
(gyro-phase averaged motion) breaks down before this moment and the
full treatment of the dynamics (including the particle phase) is required.
In particular, before the minimum of $U$$\left(r\right)$ at $r=r_{{\rm d}}$
disappears, another minimum of $U$ emerges at large $r=r_{{\rm esc}}\gg r_{{\rm d}}$.
Here $r_{{\rm esc}}$ stands for an 'escape' radius, since the particle
makes a long excursion from the filament when it moves around this
minimum of $U\left(r\right)$. In fact, it simply rotates in a weak
magnetic field and has therefore a good chance to escape into the
IGM. Naturally, a local maximum at $r=r_{{\rm s}}$ appears between
the two minima and a separatrix, which crosses the point $(r_{{\rm s}},0)$,
forms on the phase plane $\left(r,p_{r}\right)$, Fig.\ref{fig:Phase plane}
(see also Fig.\ref{fig:3Dtrajectory} for a numerical example of particle
motion). These two additional critical points emerge at $t=t_{{\rm c}}$
when

\[
\frac{\partial^{n}}{\partial r^{n}}\left(\frac{\Psi\left(r\right)-\mathcal{P}\left(t_{c}\right)}{r}\right)=0,\ \ n=1,2\]
The critical time $t_{{\rm c}}$ and the radius $r_{{\rm c}}$ are
therefore determined by the following relations:

\[
\Psi^{\prime\prime}\left(r_{{\rm c}}\right)=0;\;\;\mathcal{P}\left(t_{{\rm c}}\right)=\Psi\left(r_{{\rm c}}\right)-r_{{\rm c}}\Psi^{\prime}\left(r_{{\rm c}}\right)\]
Starting from this moment, a particle, while still oscillating and
climbing to higher energies in the left (narrow) potential well (see
Fig.\ref{fig:Potential-energy-of}) can also exercise finite motion
in the shallow right potential well, provided that particle energy
is high enough. The moment $t=t_{{\rm s}}$, and the particle momentum
$p_{{\rm s}}$ when the particle moves from the left to the right
potential well, can be determined from the following relations

\[
\mathcal{P}_{{\rm s}}=\Psi\left(r_{{\rm s}}\right)-r_{{\rm s}}\Psi^{\prime}\left(r_{{\rm s}}\right);\;\;\Psi^{\prime}\left(r_{{\rm s}}\right)=p_{{\rm s}}\]
where $\mathcal{P}_{{\rm s}}=\mathcal{P}_{0}-vt_{{\rm s}}$. These
are two equations for the three unknowns ($p_{{\rm s}},r_{{\rm s}}$
and $\mathcal{P}_{{\rm s}}$) and the conservation of adiabatic invariant
may be used as the third equation. However, when the particle orbit
approaches the separatrix, the simplified drift theory approximation
in eq.(\ref{eq:adinv}) becomes inaccurate. This is illustrated in
Figs.\ref{fig:PoftWithAdiab} and \ref{fig:Final-phase-of}. The integral
representation of the adiabatic invariant will be used below for the
treatment of this final phase. Note that this is the most efficient
phase of the acceleration mechanism. What happens is that particles,
while dwelling progressively longer at the hyperbolic point during
their oscillations in the potential well, virtually circulate around
the origin, being thus in a 'betatron' acceleration regime. The energy
gain is very fast (explosive, as we shall see) at this stage, since
the electric field of the accreting plasma is almost collinear to
the particle velocity. At the same time the orbit radius decreases
since the magnetic field increases fast enough along the particle
orbit.

\subsection{Details of particle motion\label{sub:Details-of-particle}}

To describe further details of the particle dynamics, we specify the
magnetic field profile $B\left(r\right)$. Since it is essentially
identical to the density profile (see the end of \ref{sub:Plasma-flow-towards})
we can utilize the expression in eq.(\ref{eq:Rhoasy}). For convenience,
we slightly modify it by representing $B\left(r\right)$ (in dimensionless
variables) as \begin{equation}
B\left(r\right)=r^{-\nu}+1\label{eq:Bofr}\end{equation}
with $\nu=1/\left(\gamma-1\right)$. The difference between the last
expression and the corresponding density profile is insignificant,
given the approximate character of eq.(\ref{eq:Rhoasy}). This approximation
is acceptable as it ensures the field compression that is equivalent
to the density compression when $r$ changes from $r\gtrsim1$ to
$r\ll1$.

We may now explicitly identify the separatrix parameters

\begin{equation}
\mathcal{P}_{{\rm s}}\simeq\frac{\nu-1}{2-\nu}p_{{\rm s}}^{-\mbox{\ensuremath{\left(2-\nu\right)}/\ensuremath{\left(\nu-1\right)}}}\;\;{\rm and}\;\; p_{{\rm s}}=r_{{\rm s}}+r_{{\rm s}}^{1-\nu}\simeq r_{{\rm s}}^{1-\nu},\label{eq:Psnadps}\end{equation}
that relate the particle momentum $p_{{\rm s}}$, its orbit radius
$r_{{\rm s}}$ and the canonical momentum $\mathcal{P}_{{\rm s}}$
(or, equivalently, the time $t_{{\rm s}}$) at the moment of separatrix
crossing. To simplify the calculation of the adiabatic invariant $I$
in eq.(\ref{eq:adinv}), we set the gas adiabatic index $\gamma=5/3$,
i.e. $\nu=3/2$ in eq.(\ref{eq:Bofr}). For now, it is sufficient
to evaluate the adiabatic invariant at the moment of separatrix crossing,
that means the moment when a particle goes from the left potential
well to the right one in Fig.\ref{fig:Potential-energy-of}, that
is 

\begin{equation}
I_{{\rm s}}\equiv I\left(p_{{\rm s}}\right)\approx2\intop_{r_{s,{\rm {\rm min}}}}^{r_{{\rm s}}}\sqrt{p_{{\rm s}}^{2}-\left(\frac{2\sqrt{r}-\mathcal{P}_{{\rm s}}}{r}\right)^{2}}dr\label{eq:Is}\end{equation}
where $r_{{\rm s}}\approx1/p_{{\rm s}}^{2}$, $\mathcal{P}_{{\rm s}}=1/p_{{\rm s}}$,
and $r_{{\rm s,min}}=\left(\sqrt{2}-1\right)^{2}/p_{{\rm s}}^{2}$.
A simple integration yields

\begin{equation}
I_{{\rm s}}=\frac{2}{p_{{\rm s}}}\left(4\ln\frac{1}{\sqrt{2}-1}-\pi\right)\label{eq:Is1}\end{equation}
From the conservation of adiabatic invariant we thus have

\begin{equation}
I_{0}=\pi\frac{p_{0}^{2}}{B_{0}}=I_{{\rm s}}\approx\frac{0.77}{p_{{\rm s}}}\label{eq:adinvcons}\end{equation}
where $p_{0}$ and $B_{0}$ are the particle momentum and the magnitude
of the magnetic field at a certain point in the flow far away from
the filament (where the drift approximation still applies). If a particle
of the momentum $p_{0}$ enters the acceleration process at $r_{0}\gg R_{{\rm B}}$,
then $B_{0}=1$. To include particles injected closer to the filament
axis, e.g. from a shock in the accretion flow that diffusively preaccelerates
particles, we retain the term with $B_{0}>1$. Such particles may
be injected from one of the nearby nodes (clusters) where strong shocks
are expected to diffusively accelerate cosmic rays with high efficiency,
e.g., \citep{KangCluster96,Miniati01,KangJones05,Ensslin07,Pfrommer07,Blasi04}.

\begin{figure}[h]
\includegraphics[bb=0bp 230bp 612bp 792bp,scale=0.7]{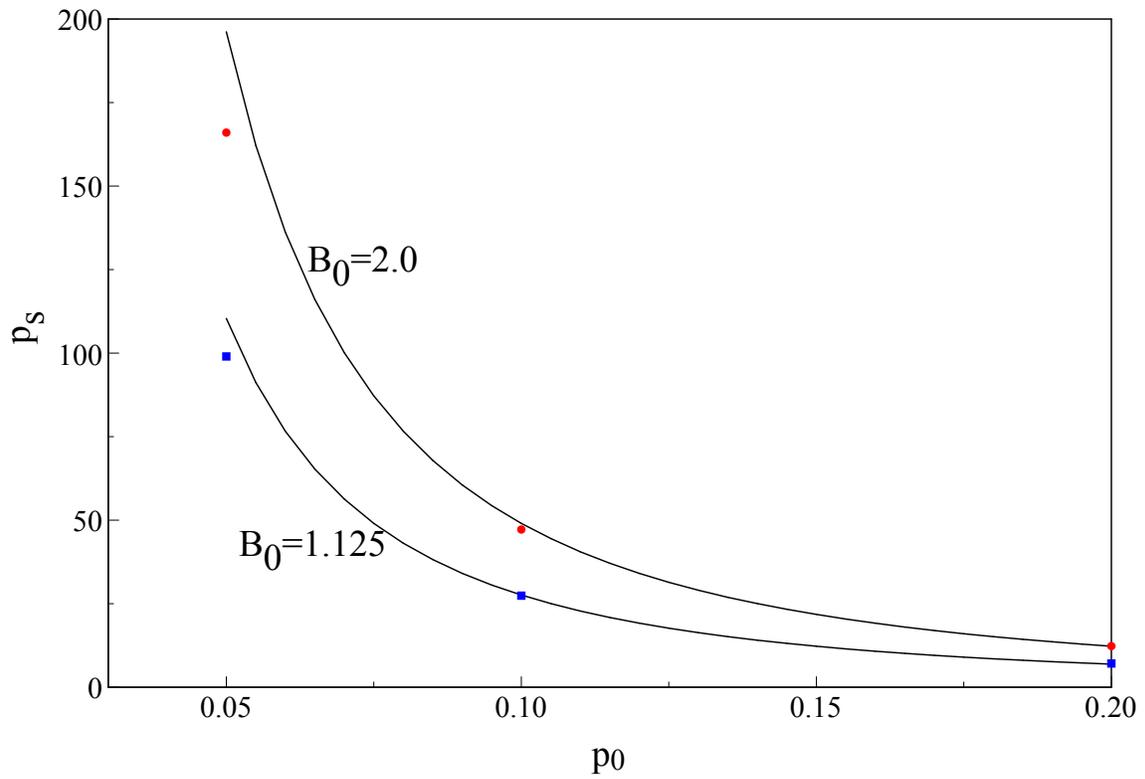}

\caption{Particle maximum momentum $p_{{\rm s}}$ (at separatrix crossing)
as a function of the initial momentum $p_{0}$, plotted for two initial
coordinates of the particle that correspond to the two indicated magnetic
field values. Solid lines represent eq.(\ref{eq:psmax}), while the
numerical solutions are shown by squares and circles. \label{fig:Particle-maximum-momentum}}

\end{figure}

From the last equation we obtain the following interesting (inverse-square)
relation between the initial and the final (separatrix value) particle
momentum

\begin{equation}
p_{{\rm s}}\approx0.25\frac{B_{0}}{p_{0}^{2}}\label{eq:psmax}\end{equation}
Fig.\ref{fig:Particle-maximum-momentum} shows the results of numerical
integration of a few particle orbits along with the curves corresponding
to eq.(\ref{eq:psmax}). Noticeable deviations are seen only at unrealistically
high values of $p_{{\rm s}}$, which are quite understandable, though
(see Sec.\ref{sub:Final-phase-of} below and the discussion in Sec.\ref{sec:Conclusions}). 

\begin{figure}[h]
\includegraphics[bb=50bp 230bp 612bp 792bp,scale=0.7]{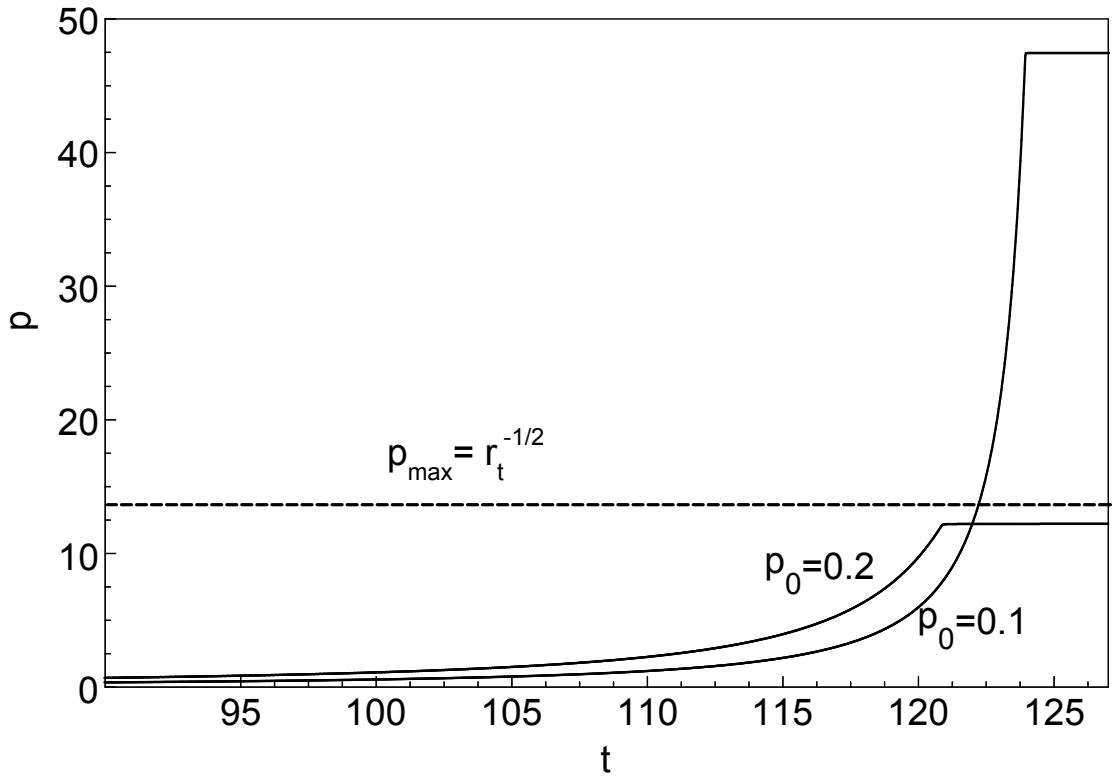}
\caption{Example of two particle orbits with different initial momenta $p_{0}=0.1,0.2$
(also shown as points in Fig.\ref{fig:Particle-maximum-momentum}).
While the trajectory with the smaller initial momentum formally ends
up at the larger final momentum, changing the flow direction towards
one of the nodes (this happens at $r=r_{t}\sim10^{-2}$, which is
marked by the dashed horizontal line) prevents the particle with the
higher final energy from escaping the filament.\label{fig:Example-of-two}}

\end{figure}

However, this seemingly strong energy gain should be taken with care.
Since $p_{{\rm s}}\approx1/\sqrt{r_{{\rm s}}}$, the maximum momentum
$p_{{\rm s}}$ is limited by the condition $r_{{\rm s}}\gtrsim r_{{\rm t}}$,
where $r_{{\rm t}}\ll1$ is the radius below which the flow changes
its direction from radial (towards filament) to axial (towards node).
Particles with sufficiently small gyro-radii are convected along the
$z$-axis out of the acceleration zone. This constitutes the re-acceleration
character of the process and constrains the initial particle momentum
$p_{0}$. In order to reach the separatrix and escape the accelerator
\emph{before} being convected towards one of its end nodes, a particle
must enter the acceleration with the momentum

\begin{equation}
p_{0}>\sqrt{0.25B_{0}}r_{{\rm t}}^{1/4}\label{eq:pinitlimit}\end{equation}
This is a significant but not the prohibiting constraint on the initial
particle momentum. The final particle momentum $p_{{\rm max}}$ is
limited by the condition $p_{{\rm max}}\le\min\left(r_{{\rm t}}^{-1/2},p_{{\rm s}}\right)$.
If $p_{{\rm s}}>r_{{\rm t}}^{-1/2}$, the particle cannot be released
from the accelerator and sinks into the filament. The illustration
of the above consideration is given in Fig.\ref{fig:Example-of-two}. 

As a proxy for $r_{{\rm t}}$, an estimate of  magnetic field and/or 
density compression between the flow outside of the accretion radius 
and the filament axis can be used. So, a $\sim10^{3}$ increase in 
$B$ from the nanogauss IGM field to the $\mu$G intracluster field 
appears to be reasonable. With the $r^{-3/2}$ scaling of the magnetic 
field adopted above, this translates into $r_{{\rm t}}\sim10^{-2}$, 
yielding, in turn, $p_{{\rm max}}\sim10$ for $p_{0}^{2}\sim1/10$. 
Therefore, if a particle enters the acceleration at a few $10^{19}$eV 
it may reach $10^{21}$eV by the moment of ejection.

Nevertheless, it is worth while to return to the dimensional variables
and recall that the length (including the Larmor radius) is measured
in the accretion radii $R_{{\rm B}}$, eq.(\ref{eq:BondiRad}), that
can be approximated as 

\begin{equation}
R_{{\rm B}}\approx2.1\frac{M_{{\rm f}}}{10^{13}M_{\sun}}\frac{10^{6}K}{T_{\infty}}{\rm Mpc,}\label{eq:RBdim}\end{equation}
where $T_{\infty}$ is the IGM temperature (i.e. well outside the
accretion radius). Thus, the maximum particle energy can be represented
as follows

\begin{equation}
E_{{\rm max}}\approx2\cdot{\rm min}\left\{ 0.25\frac{B_{0}}{B_{\infty}}\left(\frac{R_{{\rm B}}}{r_{{\rm g}0}}\right)^{2},\sqrt{\frac{R_{{\rm B}}}{r_{{\rm t}}}}\right\} \frac{M_{{\rm f}}}{10^{13}M_{\sun}}\frac{10^{6}K}{T_{\infty}}\frac{B_{\infty}}{{\rm nG}}{\rm EeV}\label{eq:Emaxdim}\end{equation}
The major uncertainty is the accretion radius, eq.(\ref{eq:RBdim}),
to which the Larmor radius and thus the maximum energy are scaled.
We note here in passing, that the accretion radius has a somewhat
ambiguous relation to the accretion rate $\dot{M}=4\pi\lim_{r\to\infty}r^{2}u_{r}\left(r\right)\rho\left(r\right)$,
as it formally involves a limiting transition if this radius is to
be determined from the ambient density and temperature. Indeed, while
the density has a definite limit $\rho_{\infty}$, the other two comprise
a $0\cdot\infty$-type uncertainty, so one is left with a sonic radius
to accept as an accretion radius $R_{{\rm B}}$. At the same time,
if $R_{{\rm B}}$ is not much larger than $l$, the form of the surface
from which the gas effectively accrets onto the filament-node 'dumbbell'
structure is more complicated than a sphere, Sec.\ref{sub:Plasma-flow-towards}.

Assuming, however, $R_{{\rm B}}>2l$, let us attempt to estimate the
absolute value of $R_{{\rm B}}$ and thus $E_{{\rm max}}$ in the
above expressions. According to the simulations of the large scale
structure formation, the gas temperature inside the filament is expected
to be in the range $10^{5}-10^{7}$K, e.g., \citep{Miniati00,Dolag06,RyuKangTW03}.
These predictions seem to be confirmed by recent observations of Abell
222/223 clusters of galaxies \citep{Abell222_08}. However, a lower
temperature may be appropriate for eq.(\ref{eq:RBdim}) as it refers
to the outskirt of the accretor, Sec.\ref{sub:Plasma-flow-towards}.
In addition, there are also observations of filaments with significantly
lower temperatures \citep{ColdFilaments06}, $T\lesssim10^{5}$K or
even $T\ll10^{5}$K, where the latter case, however, is rather an
indication of the presence of cold clouds. The mass of the Abell 222/223
filament \citep{Abell222_08} is found to be $M_{{\rm f}}\simeq10^{14}M_{\sun}$
and its length $2l\simeq15$Mpc. These figures translate into $R_{{\rm B}}$
in eq.(\ref{eq:RBdim}) that can formally reach $100$Mpc (for $T_{\infty}=2\cdot10^{5}K$)
and beyond, which would be enough to reach $10^{21}$eV for $B_{\infty}\simeq1nG$
and $\sqrt{R_{{\rm B}}/r_{{\rm t}}}\simeq10$ (see Sec.\ref{sub:Plasma-flow-towards}).
In addition, the filament end nodes may also effectively increase
$M_{{\rm f}}$ thus enhancing $R_{B}$ and $E_{{\rm max}}$ and compensating
for possibly higher $T_{\infty}$. Note that $r_{{\rm t}}$ is then
of the order of 1 Mpc (based on its simple estimates in Sec.\ref{sub:Plasma-flow-towards}
from the three orders of magnitude density and magnetic field compression
in the accreting flow) while the initial particle gyro-radius should
be of the order of $10$Mpc. 

While the observations of the cosmic web filaments are important for
the proposed acceleration mechanism, the latter is yet to be elaborated
to the level of a predictive model in which the observational parameters
may be fully used. Clearly, the size of the axial stream $r_{{\rm t}}$
remains to be consistently determined from at least two-dimensional
hydrodynamic treatment beyond the simplified approach in Sec.\ref{sub:Plasma-flow-towards}.
We  note that such treatment may reduce $r_{{\rm t}}$ thus improving
the acceleration, particularly if the gas cooling is taken into account,
e.g.,\citep{FabianCoolFlows94}. We will discuss other kinematic limitations
of the acceleration in Sec.\ref{sec:Conclusions}.

\subsection{Final phase of acceleration\label{sub:Final-phase-of}}

To complete our description of the proposed acceleration mechanism,
here we consider its final stage in some detail. This is the most
important phase of the acceleration since the energy losses, which
will be briefly addressed in the next section, become progressively
important with growing energy and may terminate the acceleration before
the required energy is reached.

As we showed in the previous section, the maximum momentum (more precisely
the momentum at the instant of separatrix crossing) may be determined
from the conservation of the particle adiabatic invariant. The particle
trajectory, including its part immediately preceding the separatrix
crossing, can be described by this principle as well. In particular,
the time dependence of the particle momentum short before the separatrix
crossing can be written as follows (see Appendix \ref{sec:Adiabatic-invariant-near})

\begin{equation}
p\left(t\right)=\frac{1}{\mathcal{P}_{0}-vt}\label{eq:PoftSep}\end{equation}
The acceleration time (conventionally defined) thus reduces towards
the end of acceleration

\begin{equation}
\tau_{a}=\frac{p}{\dot{p}}=\frac{1}{\nu p}\label{eq:tauacc}\end{equation}
which is in sharp contrast with the DSA, for example, where the acceleration
time grows linearly with $p$ (at least for Bohm diffusion). The acceleration
termination time (separatrix crossing) is somewhat uncertain, because
the separatrix itself is determined for a fixed value of $\mathcal{P}$.
In reality, $\mathcal{P}$ slowly changes in time, which leads to
the separatrix deformation and saddle point displacement, a phenomenon
closely related to the separatrix splitting under perturbations in
dynamical systems. In simple terms it can be understood here by observing
that if a particle is about to cross the separatrix almost exactly
through the hyperbolic point, it can still make an extra cycle by
an infinitesimal (e.g. numerical round-off errors) orbit perturbation.
The extra cycle will then have a finite ($\sim v\ln v$) effect on
the particle dynamics. The effect is pronounced, since the period
diverges when the particle approaches the separatrix:

\begin{equation}
T=\frac{\partial I}{\partial p}\simeq\frac{2\sqrt{2}}{p_{{\rm s}}^{2}}\ln\frac{1}{1-p\mathcal{P}}.\label{eq:Tperiod}\end{equation}
Here $1-p\mathcal{P}\ll1$ can be expressed through the time $t$
using eqs.(\ref{eq:eps}) and (\ref{eq:pOfP}). This phenomenon is
predominantly responsible for the discrepancies between the analytic
and numerical calculations shown in Fig.\ref{fig:Particle-maximum-momentum}.
Further discussion can be found in Appendix \ref{sec:Adiabatic-invariant-near}
and in Sec.\ref{sec:Conclusions}.

\section{Energy losses and other energetic constraints\label{sec:Losses}}

Often, it is not the maximum energy of an accelerator which precludes
the proton production with $E\gtrsim10^{20}$eV but the losses that
either the accelerator cannot compensate for or exit losses (i.e.
caused by strong photon and/or magnetic fields, surrounding the acceleration
zone) \citep{NormanAchterb95,Levinson00,JonesUHECR04}. The DSA for
example, is relatively slow because of the long idling time and, more
importantly, it slows down with growing energy. Although the acceleration
time may level off under certain circumstances, e.g. \citep{MD06},
it is unlikely to decrease with $E$ and thus, the Synchrotron-Compton
losses, for example, having the loss rate $\propto E$ (so that $\dot{E}\propto-E^{2}$)
will ultimately prevail. Aside from that, the operation of powerful
accelerators is usually accompanied by the production of other forms
of energy, such as magnetic and photon fields, that boost the Synchrotron-Compton
losses.

Fortunately, the present acceleration mechanism rapidly speeds up
towards the maximum energy, $\dot{p}=vp^{2}$, eq.(\ref{eq:PoftSep}).
The maximum energy is reached when a particle crosses the separatrix
at $p_{{\rm max}}=p_{{\rm s}}$. Although the energy grows also beyond
this point, its growth is very slow as the particle recedes from the
filament so that the separatrix crossing may be considered as an abrupt
end of the acceleration. It is extremely beneficial for the UHECR
production to terminate the acceleration process in such a way; the
synchrotron losses also drop abruptly as the particle is released
into a void, where only a weak magnetic field is present.

The Synchrotron-Compton losses of a proton with an energy $E$ can
be written as

\[
\dot{E}_{{\rm sc}}=-\frac{2}{3}\frac{e^{4}}{m_{{\rm p}}^{4}c^{7}}B^{2}E^{2}\]
where $e$ and $m_{{\rm p}}$ denote the proton charge and mass. Using
dimensionless variables introduced in Sec.\ref{sec:Acceleration},
we can add this term to eq.(\ref{eq:pdot-1}) as follows

\begin{equation}
\dot{p}=\frac{vp_{\vartheta}}{rp}-\eta p^{2}B^{2}.\label{eq:pdotloss}\end{equation}
Here the dimensionless loss rate is

\[
\eta=\frac{2}{3}\left(\frac{m_{{\rm e}}}{m_{{\rm p}}}\right)^{4}\frac{r_{e}R_{{\rm B}}^{2}}{r_{{\rm g}\infty}^{3}}\approx3\cdot10^{-12}\left(\frac{R_{{\rm B}}}{10{\rm Mpc}}\right)^{2}\left(\frac{B_{\infty}}{{\rm nG}}\right)^{3},\]
with the classical electron radius $r_{e}=e^{2}/m_{{\rm e}}c^{2}$,
the Bondi radius $R_{{\rm B}}$, eq.(\ref{eq:BondiRad}), and $r_{{\rm g}\infty}=m_{{\rm e}}c^{2}/eB_{\infty}$.
Naturally, the losses may become important towards the end of acceleration
as both $p$ and $B$ grow. As we know the acceleration term is close
to $\dot{p}=vp^{2}$ at this point, eq.(\ref{eq:tauacc}), so that
from eq.(\ref{eq:pdotloss}) we find that the acceleration will proceed
up to the maximum momentum $p=p_{{\rm s}}$ if 

\begin{equation}
\frac{B^{2}\left(r_{{\rm s}}\right)}{B_{\infty}^{2}}<\frac{v}{\eta}\simeq5\cdot10^{7}\sqrt{\frac{T}{10^{4}{\rm K}}}\left(\frac{B_{\infty}}{{\rm {\rm nG}}}\right)^{-3}\left(\frac{R_{{\rm B}}}{{\rm 10{\rm Mpc}}}\right)^{-2}\label{eq:syncomp}\end{equation}
Here we have evaluated the parameter $v$, eq.(\ref{eq:vpar}), by
simply taking it at the sonic point (and using the approximation of
the flow and field compression between $r=\infty$ and $r=R_{{\rm B}}$,
(eq.{[}\ref{eq:rhotorhoing}{]} ) as $v\simeq4C_{\infty}/c$. Moreover,
if we use our earlier estimate of the IGM magnetic field $\sim{\rm nG}$
compressed to $\sim\mu{\rm G}$ level, for example, the l.h.s. of
the last inequality is $\sim10^{6}$ and even a strong inequality
in eq.(\ref{eq:syncomp}) can be satisfied. Of course, $R_{{\rm B}}$
can be estimated as $R_{{\rm B}}\sim GM_{{\rm f}}/C_{{\rm s}}^{2}$
but it can too be inferred from the simulations and used then in eq.(\ref{eq:syncomp}).

Turning to the photo-pion losses on the CMB we note that, in contrast
to the Synchrotron-Compton losses, it is essentially a threshold process.
Moreover, the photo-pion losses strongly dominate the pair production
losses above a few $10^{19}$eV, so we may ignore the latter in the
energy range of our interest. In what follows, we will extensively
utilize the analyses of \citet{BerezBook90,BerezGrig88,NormanAchterb95}.
In particular, for energies $E<3\cdot10^{20}$eV the following simplified
representation of the loss term can be used

\begin{equation}
-\frac{\dot{E}}{E}\approx\frac{c}{l_{\pi}}e^{-E_{{\rm th}}/E}\label{eq:edotovere}\end{equation}
with $E_{{\rm th}}=m_{{\rm p}}m_{\pi}c^{4}\left(1+m_{\pi}/m_{p}\right)/2kT\approx3\cdot10^{20}$eV
and $l_{\pi}\simeq10$ Mpc. Here $m_{\pi}$ is the pion's rest mass
and $T$ is the $2.7$K CMB radiation temperature. For higher energies
$E\gg E_{{\rm th}}$ a slightly lower value of $c/l_{\pi}\approx1.8\cdot10^{-8}$yr$^{-1}$
may be adopted but, this energy range can hardly be reached by this
acceleration mechanism.

The dimensionless acceleration rate at the final stage of acceleration
can be written as $p^{-1}\dot{p}\simeq vp$, eq.(\ref{eq:tauacc}).
Combining this with eq.(\ref{eq:edotovere}) and using the dimensionless
variables, we obtain

\begin{equation}
\frac{\dot{p}}{p}=vp-\frac{R_{{\rm B}}}{l_{\pi}}e^{-p_{{\rm th}}/p}\label{eq:pdotpion}\end{equation}
where $p_{{\rm th}}$ is the dimensionless photo-pion threshold momentum
$E_{{\rm th}}/c$. The right hand side of this equation may either
have two roots or none. Recalling that the particle gyroradius is
normalized to $R_{{\rm B}}$, we may write the condition for the latter
case ($\dot{p}/p>0$ for all $p$) as follows

\[
v\frac{r_{{\rm g}}\left(p_{{\rm th}}\right)l_{\pi}}{R_{{\rm B}}^{2}}>e^{-1}\approx0.37\]
where $r_{{\rm g}}\left(p_{{\rm th}}\right)$ is the gyroradius of
a proton with the momentum $p=p_{{\rm th}}$ in the $B_{\infty}$
magnetic field. Using our estimate $v\simeq4C_{\infty}/c$ once again,
the last condition rewrites

\begin{equation}
\frac{C_{\infty}}{c}>3\cdot10^{-3}\left(\frac{R_{{\rm B}}}{{\rm 10Mpc}}\right)^{2}\left(\frac{B_{\infty}}{{\rm nG}}\right).\label{eq:photlim}\end{equation}
This requirement does not seem to be totally unrealistic particularly
for flows with $T\sim10^{6}-10^{7}K$ heated by strong external shocks
in the structure formation \citep{Miniati00,CenOstriker06}. Most
likely it is marginally violated, so that the two roots of the expression
on the r.h.s of eq.(\ref{eq:pdotpion}) do exist and significant
photo-pion losses occur between these energies. It should also be
recognized that higher maximum energy in eq.(\ref{eq:Emaxdim}) makes
the inequality in eq.(\ref{eq:photlim}) more difficult to satisfy
in terms of parameters, such as $C_{\infty}$ and $B_{\infty}$. The
environments where the both requirements can be met should not be
very common.

Overall, due to the fast energy gain in the betatron acceleration
phase, the energy losses that are fatal for the DSA may be overcome.
We therefore conclude that the maximum energy is likely to be determined
by intrinsic limitations of the acceleration mechanism and not by
the energy losses. The intrinsic acceleration limit is set by either
the separatrix crossing or by reaching the flow deflection inside the filament,
whichever occurs first. 

Even though the calculation of the shape of the spectrum is beyond
the scope of this paper, the key points of such calculation are worth
mentioning here. First of all, the inverse-square relation between
the initial and final particle momentum, eq.(\ref{eq:psmax}), suggests
flipping the injected spectrum with respect to the fixed point of
the map $p_{0}\mapsto p_{{\rm s}}$, given by (eq.{[}\ref{eq:psmax}{]}).
Obviously, the fixed point of the map is $p=\left(0.25B_{0}\right)^{1/3}$.
If the injection spectrum has the form $f_{{\rm inj}}\propto p^{-q}$,
for example, and it should be taken in the interval $0.5\sqrt{B_{0}}r_{{\rm t}}^{1/4}<p<\left(0.25B_{0}\right)^{1/3}$,
then the accelerated particle spectrum will cover the interval $\left(0.25B_{0}\right)^{1/3}<p<r_{{\rm t}}^{-1/2}=p_{{\rm max}}$,
with an index $q^{\prime}=q/2.$ However, at least three obvious phenomena
can steepen the spectrum substantially. First, as the particle momentum
approaches $p_{{\rm max}}$, its orbit crosses and recrosses the circle
of the radius $r_{{\rm t}}$, and the odds for particle convection
with the flow towards one of the nodes progressively increase. Second,
the boundary between the radial and axial accretion at $r=r_{{\rm t}}$
is not sharp. Therefore, one should not expect a sharp spectrum cut-off
at $p_{{\rm max}}$ but rather its decline starting at lower momenta.
Finally, the photo pion losses that should be relatively strong for
particles with momenta between the two roots of the r.h.s. of eq.(\ref{eq:pdotpion})
will modify the spectrum substantially. These phenomena may easily
compensate for the spectrum hardening, produced by the inverse-square
relation between initial and final momenta. Nevertheless, the kink
at $p=\left(0.25B_{0}\right)^{1/3}$ may be pronounced. It should
be noted here that energy-losses, such as the pair production losses
and photo pion losses, have already been suggested to be responsible
for such universal features in the UHECR energy spectrum as the dip
and the bump \citep{Berez06}.

A useful test of accelerator's potential is the energy dissipation
rate by the emf (electromotive force) required for the acceleration
of particles to the energy $E_{{\rm max}}$, subjected to the vacuum
impedance $Z_{0}\sim10^{2}\Omega$ \citep[e.g.,][]{Bland00,Sigl09}.
Specifically, this power is $L_{{\rm min}}\sim\mathcal{E}^{2}/Z_{0}$,
with the induction in the case of our interest $\mathcal{E=}$2$\pi rE_{\vartheta}=2\pi ruB/c=2\pi vB_{\infty}R_{{\rm B}}$.
This quantity, in contrast to the most acceleration schemes, where
$\mathcal{E}\sim E_{{\rm max}}/e$, is not related to $E_{{\rm max}}$
because $E_{{\rm max}}$ is independent of the accelerator power $v$,
Sec.\ref{sub:Details-of-particle}. The reason for this is that in
the betatron regime particles may pass through the accelerating field
multiple times, and they will do it as many times as needed to reach
the maximum energy. Indeed, according to eq.(\ref{eq:pthetarelation})
we may write the dimensionless $p_{{\rm max}}\approx2\pi nv$, where
$n$ is the number of rotations around the filament (orbit winding
number), needed to reach $p_{{\rm max}}$. Therefore, returning to
the dimensional units we can represent the induction also as $\mathcal{E}\sim E_{{\rm max}}/en$,
and $L_{{\rm min}}$ is thus reduced by $n^{2}$, $L_{{\rm min}}\sim E_{{\rm max}}^{2}/e^{2}n^{2}Z_{0}$.
On the other hand, the acceleration time does depend on the power
$v$ and a low-power source may fail to accelerate particles to the
required energy for the energy loss reason.

\section{Summary and Outlook\label{sec:Conclusions}}

We have suggested a mechanism capable of proton re-acceleration to
the maximum energy $\sim10^{21}$eV, provided that particles with
energies $\gtrsim10^{19}$eV are seeded. These encouraging figures
emerge for the plasma accretion on to a dark matter filament with
the magnetic field compressed by a factor $\sim10^{3}$, say from
nG intergalactic field to $\mu{\rm G}$ intracluster field. The required
seed particles can arguably be pre-accelerated up to $\gtrsim10^{19.5}$eV
by the standard diffusive shock acceleration (DSA) mechanism in the
structure formation shocks \citep{NormanAchterb95,KangCluster96,JonesUHECR04}
within the same accretion flow. Another important advantage of the
suggested mechanism over, e.g., the DSA, is its very high rate during
the end phase of acceleration, when particles usually suffer catastrophic
losses. 

The calculation of the spectrum of accelerated particles and acceleration
efficiency is deferred to a future study. However, some of the ignored
phenomena, such as magnetic field perturbations, merit a brief discussion.
The magnetic field perturbations can scatter particles into the loss
cones of the magnetic mirrors, presumably supported by the accreting
nodes at the ends of the filament. However, since the final stage
of acceleration (roughly the top decade in energy) is very short,
relaxing our assumption $p_{\parallel}\ll p$ will not result in significant
particle losses along the filament before they are expelled radially.
More serious are possible losses of seed particles during the slow
drift phase between the node and filament radii, $R_{{\rm f}}<r<R_{{\rm n}}$,
where $R_{{\rm n}}\gtrsim R_{{\rm {\rm f}}}$ (Sec.\ref{sec:Accretion-flow}).
While drifting towards the filament, the particles may diffuse to
the nodes and disappear there without replenishment. The requirement
on the radial flow velocity to avoid these losses is $u/c>R_{{\rm n}}r_{{\rm g}}/l^{2}$,
where we have (somewhat arbitrarily) assumed the Bohm diffusion along
the filament. Assuming also the gyroradius at the entrance $r_{{\rm g}}\ll R_{{\rm n}}\ll l$,
the above condition does not seem to be unrealistic. Moreover, the
loss of particles to the nodes may be compensated by the injection
of high-energy particles accelerated in the nodes by, e.g., the DSA
mechanism. Appart from the axial transport, the magnetic perturbations
can result in the radial transport. The latter should also have a
twofold impact on the acceleration. It can result in a premature expulsion
of some particles from the accelerator but it can also prolongate
the betatron phase of some other particles, thus increasing the maximum
momentum.

One more factor which influences the duration of the betatron phase,
is a sharply growing ($\propto r^{-\nu}$, with $\nu>1$) magnetic
field at small $r$. This is justified when a particle is expelled
from the filament vicinity before reaching the filament axis where
the magnetic field changes its $r$- dependence to remain finite at
the origin. In terms of the dynamical system given by eqs.(\ref{eq:prham}-\ref{eq:Hamilt}),
the saddle point $r=r_{{\rm s}}$ corresponds to an unstable periodic
(circular) orbit of the full system of eqs(\ref{eq:prn}-\ref{eq:thetadot})
with $v=0$. If the magnetic field decays slower than $1/r$ ($\nu<1$)
at the fixed point $r=r_{{\rm s}}$, this circular orbit becomes stable.
We have not considered particle acceleration in this regime for the
reason that such particles are likely to be removed from the acceleration
zone with the flow along the filament, that is by the the process
not studied in the paper. On the other hand, the change of stability
of the fixed point should also retain orbits and enhance the energy.

Perhaps more important is the possible impact of perturbations on
the motion near the separatrix, which is the most productive phase
of this acceleration scheme. Recall that the particle orbit contracts
to the filament axis during the betatron phase of acceleration. Clearly,
this cannot continue infinitely so that upon approaching the separatrix,
and the unstable hyperbolic point in particular, the particle slings
out of the filament with all the energy gained during the orbit contraction.
However, there is a phenomenon of stochastic layer formation in phase
space near a separatrix. It results from perturbed particle dynamics
closely related to the famous Poincare's separatrix splitting, e.g.,
\citep{ZaslavWeakChaos91}. In general, the stochastic particle motion
near the separatrix renders the particle exit moment, and thus the
maximum energy, only statistically predictable, as opposed to our
deterministic treatment in Sec.\ref{sec:Acceleration} (see particularly
the end of Sec.\ref{sub:Final-phase-of} and Fig.\ref{fig:Particle-maximum-momentum}).
Note that the magnetic field perturbations, that can easily randomize
particle trajectory, do not have to be unstable or turbulent. The
first azimuthal mode, that is carried over (even though suppressed)
from an ambient magnetic field which is inclined with respect to the
filament, but otherwise perfectly homogeneous, \ref{sub:Magnetic-field-around},
would suffice to create the stochastic layer.

We may conclude from the last remark that the mechanism requires a
good magnetic field alignment with the filament direction. Despite
the ubiquity of filaments in the large scale cosmic structure, only
a small fraction of them may qualify for an efficient and visible
accelerator. This needs to be taken into account when considering
the observed correlation of the highest energy CRs with the distribution
of the large-scale structure. In addition to the field alignment,
the filament should be close to the plane of the sky, as the direction
perpendicular to the filament is preferable for the acceleration.
Unfortunately, for simple geometry reasons the filaments are more
easily observed (as X-ray emitting gas) in the opposite case of the
line-of-sight alignment, e.g. \citep{Abell222_08}. A more detailed
phenomenology discussion would be too speculative at this stage, but
given the ongoing debates on the UHECR arrival anisotropy, e.g., \citep{George08,Ghisellini08,Gorbunov08,Waxman09,Fargion10,Gureev10,Semikoz10},
the obvious angular characteristics of the suggested filament accelerator
are worth mentioning. Evindently, a monoenergetic particle beam should
fill a line of the angular length $\sim L/D$ in the filament direction,
where $L\ll D$ is the length of the filament and $D$ is the distance
to the observer. In the perpendicular direction (orbit plane), the
arrival direction should make an energy dependent angle $\alpha$
with respect to the line of sight to the filament, $\sin\alpha\simeq\alpha\simeq2\left(R_{{\rm B}}/r_{{\rm g}}\right)\left[\sqrt{R_{{\rm B}}/D}+D/4R_{{\rm B}}\right]$,
where the accretion radius is assumed to be much smaller than the
particle gyroradius $R_{B}\ll r_{{\rm g}}$. The above relation for
$\alpha$ may be obtained from the conservation of particle canonical
momentum $\mathcal{P}=\mathcal{P}_{{\rm s}}$ after crossing the separatrix,
Sec.\ref{sec:Acceleration}.

Overall, a particle acceleration by this mechanism to the up to date
record energy of a few $10^{20}$eV and its expulsion into the observer's
direction may be exceptional but possible. The above constraints,
however, may be useful in explaining the Centaurus event excess under
the absence of the Virgo events.

\appendix

\section{Details of the flow and magnetic field structure\label{sec:ApFlowAndMF}}

\subsection{Dark matter gravitational potential\label{sub:Dark-matter-gravitational}}

To describe the plasma accretion onto a filamentary structure outlined
in Sec.\ref{sec:Accretion-flow}, we need to specify the gravitational
potential of this structure. As we already mentioned, the DM particles
interact collectively through the gravitational potential $\Phi$
that is governed by the Poisson equation

\begin{equation}
\Delta\Phi=4\pi G\rho_{{\rm DM}}\label{eq:Poiss}\end{equation}
where $\rho_{{\rm DM}}$ is the DM mass density. Furthermore, we assume
that the system has already reached a {}``quasi-equilibrium'' in
terms of an appropriate coarsegraining \citep{LyndenB67,WhiteRees78,Navarro97}
and this self-gravitating system may be described by a stationary
Vlasov equation. The isotropic in velocity space solution amounts
to a hydrostatic equilibrium 

\begin{equation}
\nabla P_{{\rm DM}}=-\rho_{{\rm DM}}\nabla\Phi\label{eq:hydrost}\end{equation}
where $P_{{\rm DM}}$ is the DM pressure. Various approaches to the
closure problem of the above equations have been suggested in the
literature, including the recent unified DM model \citep[e.g.,][]{ScherrerUDM04}.
For our purposes it is, perhaps, sufficient to assume a simple polytrope
$P_{{\rm DM}}=P_{0}\left(\rho_{{\rm DM}}/\rho_{0}\right)^{\gamma_{{\rm DM}}}$,
where index $0$ refers to the pressure and density values at the
origin. Normalizing the spatial scale to 

\begin{equation}
L=\frac{C_{{\rm DM}}}{\sqrt{4\pi G\left(\gamma_{{\rm DM}}-1\right)\rho_{0}}},\label{eq:DMscale}\end{equation}
where $C_{{\rm DM}}^{2}=\gamma_{{\rm DM}}P_{0}/\rho_{0}$, and introducing
$\theta^{n}=\rho_{{\rm DM}}/\rho_{0}$ with $n=1/\left(\gamma_{{\rm DM}}-1\right)$,
we combine eqs.(\ref{eq:Poiss}) and (\ref{eq:hydrost}) into the
following Emden-Fowler equation

\begin{equation}
\frac{1}{r^{d-1}}\frac{\partial}{\partial r}r^{d-1}\frac{\partial\theta}{\partial r}=-\theta^{n}\label{eq:EmdFowl}\end{equation}
where $d=2,3$ for cylindrically and spherically symmetric cases of
our interest, both subjected to the boundary conditions $\theta\left(0\right)=1,\;\;\theta^{\prime}\left(0\right)=0$.
This equation was studied long time ago, including the case of self-gravitating
cylinder \citep{Ostrik64}, where a series in power of $r$ was obtained
for an arbitrary $n$. 

In choosing the suitable value for the index $n$, the superset Vlasov-Boltzmann
equation for the DM distribution may be used

\begin{equation}
\frac{\partial f}{\partial t}+\mathbf{v}\cdot\nabla f-\nabla\Phi\cdot\frac{\partial f}{\partial\mathbf{v}}=0\label{eq:Vlasov}\end{equation}

The {}``quasi-equilibrium'' solution should depend on the energy
integral $\epsilon=v^{2}/2+\Phi$. Note that we assume the independence
of the solution of the angular momentum (e.g., \citep{UDMpolytr09}).
Assuming an equipartition in energy that runs from the bottom of the
potential well $\epsilon_{min}=\min_{r}\Phi\left(r\right)$ to $\epsilon_{max}<0$,
it is straightforward to show that in the 3D case (gravitating node),
the relation between the DM density and pressure is $P_{{\rm DM}}\propto\rho_{{\rm DM}}^{5/3}$,
thus implying $\gamma_{{\rm DM}}=5/3$ and $n=3/2$. In 2D (cylindrical
symmetry) one obtains $\gamma_{{\rm DM}}=2$, $n=1.$ The latter case
is particularly simple as the solution of eq.(\ref{eq:EmdFowl}) is
given by a Bessel function $J_{0}\left(r\right)$ for $r<j_{01}\simeq2.4$,
$J_{0}\left(j_{01}\right)=0$. 

\begin{figure}[h]
\includegraphics[bb=0bp 180bp 792bp 612bp,scale=0.7]{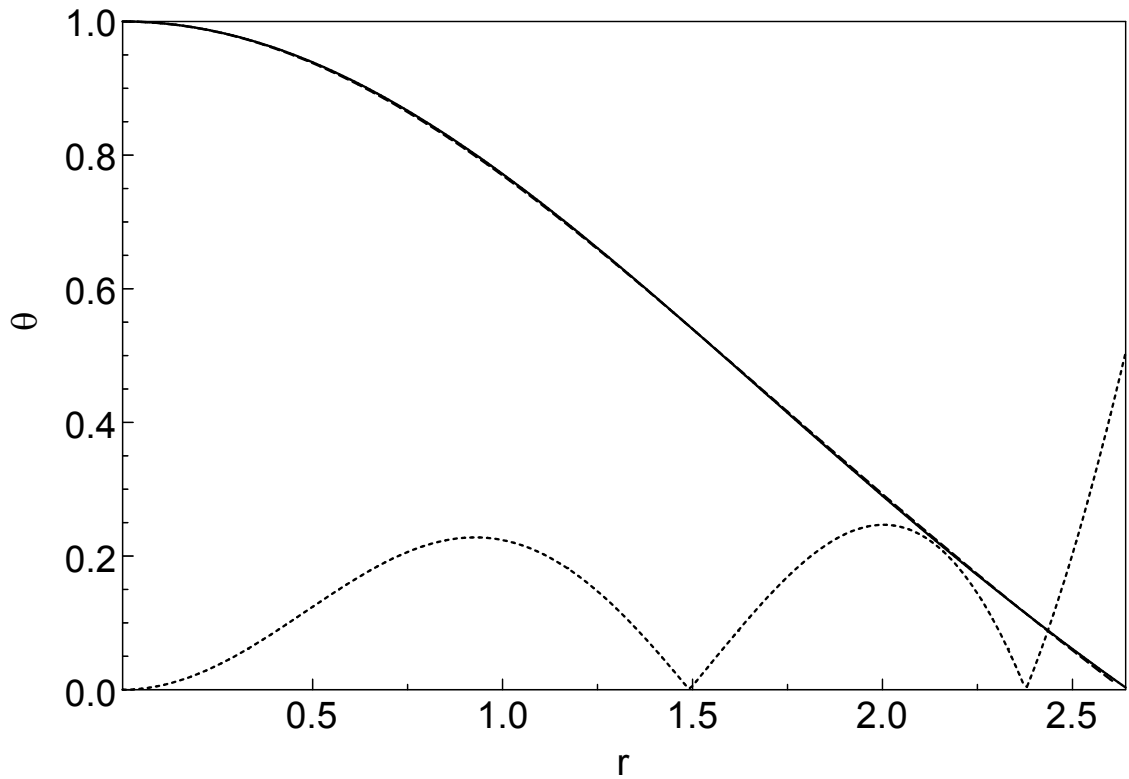}
\caption{Numerical solution of eq.(\ref{eq:EmdFowl}) ($d=2$, cylindrical
symmetry) and its Pad$\acute{{\rm e}}$ approximation, plotted as
the solid and the dashed line, respectively. As the difference is
barely visible, the absolute value of it multiplied by $100$ is shown
with the dotted line. The solution for $d=3$ is similar, and therefore
not shown here.\label{fig:Pade2D}}
 
\end{figure}

However, as the nature of DM is unknown, we also apply for comparison
the conventional $\gamma_{{\rm DM}}=5/3$ polytrope to the case of
cylindrical symmetry. Note that a closed form solution of eq.(\ref{eq:EmdFowl})
is not possible in the case $n=3/2$, since the solution clearly has
a movable pole of the form $\theta\propto400\left(r-a\right)^{-4}$,
where $a$ is a constant. The pole and the zero of $\theta\left(r\right)$
at finite $r$ (compact mass distribution), suggest the following
Pad$\acute{{\rm e}}$ approximant (for the both symmetries)

\begin{equation}
\theta=\frac{1-r^{2}/a_{d}^{2}}{1+r^{2}/b_{d}^{2}}\label{eq:Pade}\end{equation}
For $a_{2}\approx2.64$, $b_{2}\approx2.99$ (gravitating cylinder)
and $a_{3}\approx3.6$, $b_{3}\approx3.2$ (gravitating sphere), the
above approximation accurately reproduces the solution of eq.(\ref{eq:EmdFowl})
for $n=3/2$. It is shown in Fig.\ref{fig:Pade2D} along with the
numerical solution of eq.(\ref{eq:EmdFowl}). We note that there is
no significant difference between the $n=3/2$ and $n=1$ cases, where
the latter may also be appropriate for the cylindrical symmetry.

Given the filament mass $M_{{\rm f}}$, from eqs.(\ref{eq:Poiss},\ref{eq:EmdFowl})
we obtain the following expression for $C_{{\rm DM}}$ in eq.(\ref{eq:DMscale}): 

\[
C_{{\rm DM}}^{2}=\mbox{\ensuremath{\left(\gamma_{{\rm DM}}-1\right)}}\frac{GM_{{\rm f}}}{\theta_{*}l}\]
where $\theta_{*}=-a_{2}\theta^{\prime}\left(a_{2}\right)\approx1.12$,
and $2l$ is the filament length.

\begin{figure}[h]
\includegraphics[bb=150bp 0bp 620bp 540bp,scale=0.8]{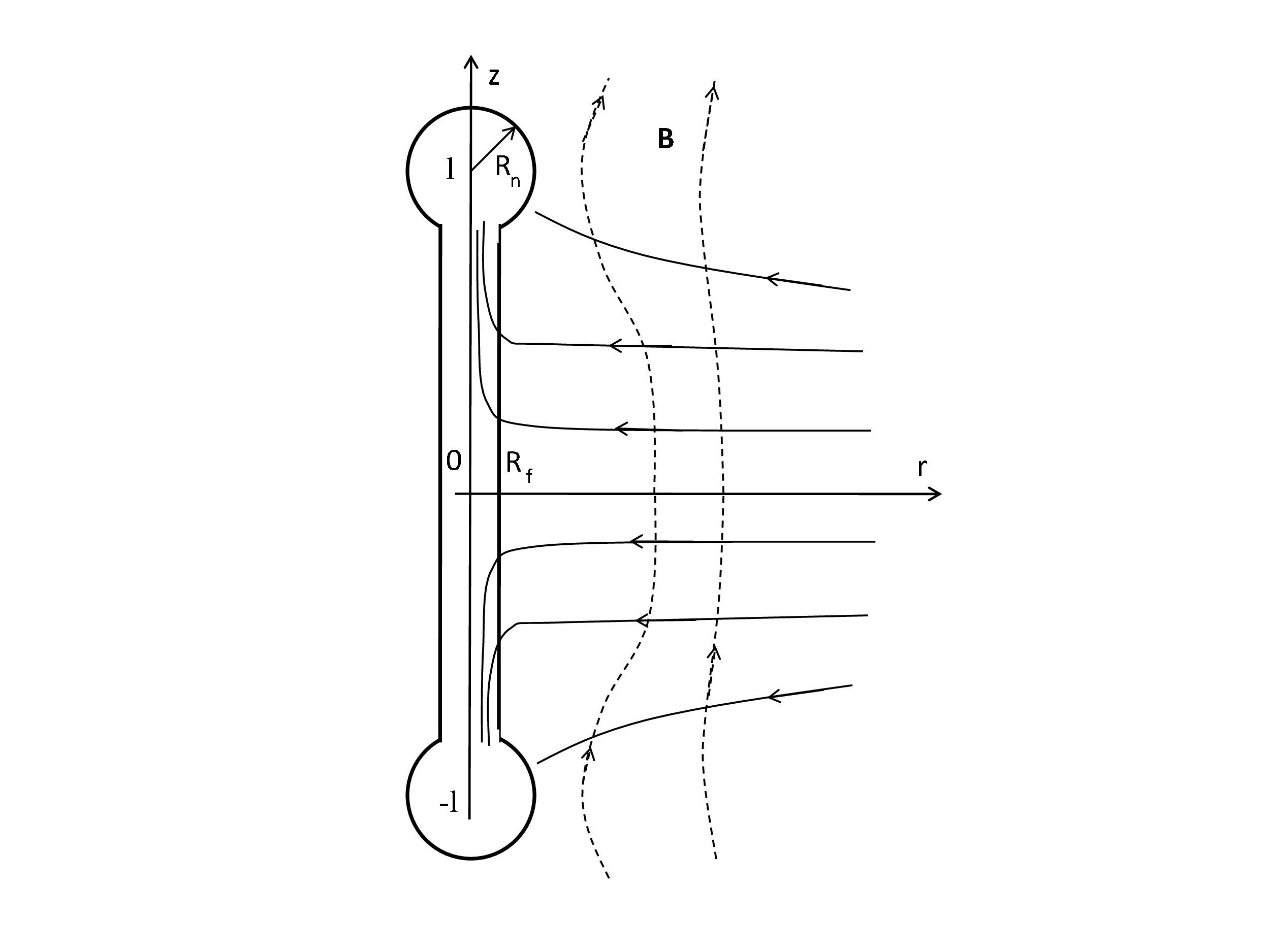}

\caption{Sketch of the flow pattern (solid lines with arrows) and magnetic
field geometry (dashed lines) near the 'dumbbell' structure consisting
of one filament and two nodes at its ends.\label{fig:Dumbbell}}

\end{figure}

It should be clear that the 'dumbbell' structure, shown in Fig.\ref{fig:Dumbbell},
can be described by combining the spherical and cylindrical solutions,
given by eq.(\ref{eq:Pade}), only approximately. Note that $a_{2}$
is identified with $R_{{\rm f}}$ - the filament radius and $a_{3}$
with the $R_{n}$ - the node radius. At a minimum, a transition region
between the two solutions should be addressed to match them. Even
if the compound solution is constructed, the 'dumbbell' cannot be
in exact equilibrium as an isolated self-gravitating structure. However,
as a part of larger assembly of similar structures, the dumbbell may
be considered to be stable. 

With the above reservations in mind and, being interested primarily
in the behavior of the gravitational potential in the middle part
of the filament (i.e. not too close to the nodes, but with a reasonably
accurate description of also the radially remote part of the solution,
$r\gg l$), the gravitational potential of an isolated 'dumbbell'
can be written down as follows (we use the cylindrical coordinates
with a $z$- axis along the filament, and with the node centers at
$z=\pm l$, Fig.\ref{fig:Dumbbell}):

\begin{equation}
\Phi\left(r,z\right)=\left\{ \begin{array}{cc}
-\frac{GM_{{\rm f}}}{\theta_{*}l}\theta\left(r\right)+\Phi_{{\rm e}}\left(R_{{\rm f}},z\right), & \;\; r<R_{{\rm f}},\;\left|z\right|<l-R_{{\rm n}}\\
\Phi_{{\rm e}}\left(r,z\right), & \;\; r\ge R_{{\rm f}}\end{array}\right.\label{eq:gravpot}\end{equation}
Here the function $\theta\left(r\right)$ and parameter $\theta_{*}$
may be chosen according to either $n=3/2$ or $n=1$ DM model, as
discussed above. Quantitatively, the difference between them is not
significant, so it is sufficient to adopt $n=3/2$ value, for example.
Furthermore, the function $\Phi_{{\rm e}}$ (specified below) which
represents the gravitational potential outside of the filament, i.e.
in the region where significant part of particle energy is gained,
does not depend on the exact distribution of the gravitational potential
inside the filament. Since $\Phi_{{\rm e}}$ represents the overall
$\Phi\left(r,z\right)$ outside of the filament edge at $r=R_{{\rm f}}$,
and since $\theta\left(R_{{\rm f}}\right)=0$, the function $\Phi_{{\rm e}}$
can be written down as follows

\begin{equation}
\Phi_{{\rm e}}\left(r,z\right)=-\frac{GM_{{\rm f}}}{2l}\sum_{\pm}\sinh^{-1}\left(\frac{l\pm z}{r}\right)-G\sum_{\pm}\frac{M_{{\rm n}}}{\sqrt{r^{2}+\left(z\pm l\right)^{2}}}\label{eq:gravpotext}\end{equation}
Here the first term is the gravitational potential of the filament
with the total mass $M_{{\rm f}}$ and length $2l$, while the second
term represents the contribution of two nodes of equal masses $M_{{\rm n}}$
at the both ends of the filament. What is important for our present
purposes, is the behavior of $\Phi$ near the mid-plane $z=0,$ inasmuch
we will consider particle acceleration primarily in this area. So,
in what follows we  set $z=0$ in the last equation, end neglect the
gravitational pull from the both nodes, for simplicity.

\subsection{Plasma flow towards filament\label{sub:Plasma-flow-towards}}

\begin{figure}[h]
\includegraphics[bb=0bp 130bp 792bp 612bp,scale=0.7]{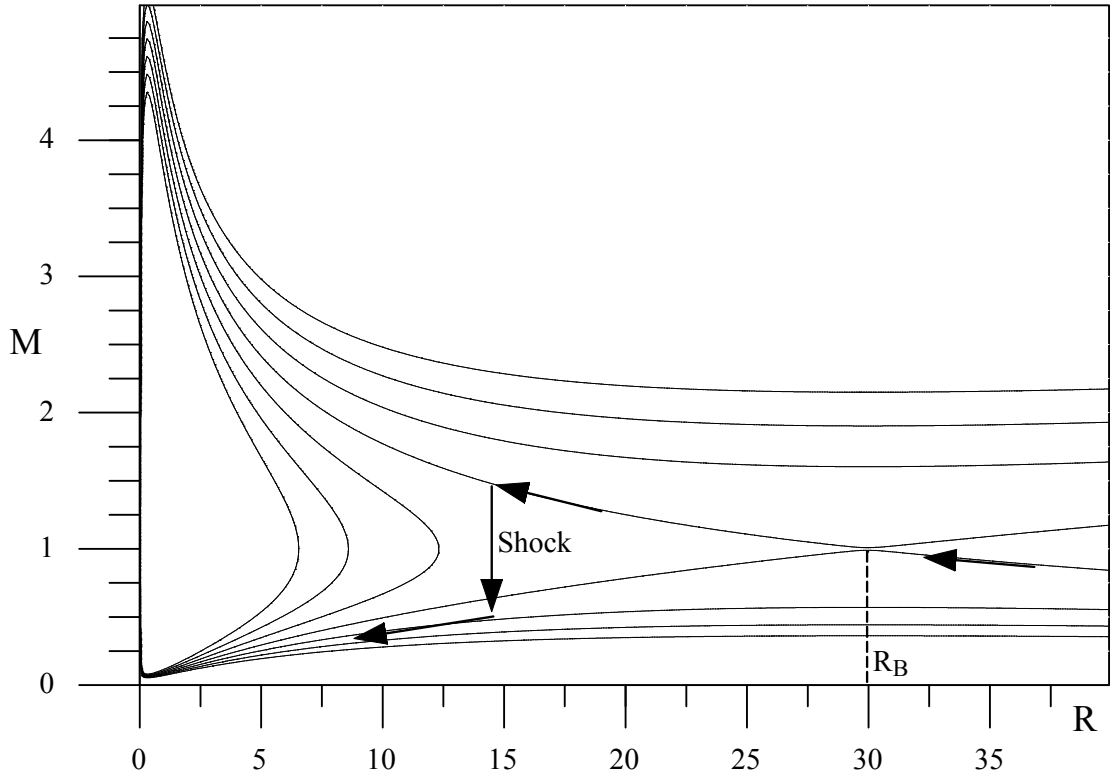}

\caption{Bondi diagram for a radial accretion flow onto a filament shown in
distance $R=r/l$ - Mach number $\mathcal{M}$ variables. A gas envelope
(arrows) passes through a sonic point at $\mathcal{M}=1$, $R=R_{{\rm {\rm B}}}$.
Then, after being accelerated and shocked, it jumps to a lower branch
(vertical arrow) of the flow diagram, decelerates further and becomes
strongly compressed before turning towards one of the nodes, shown
in Fig.\ref{fig:Dumbbell}.\label{fig:Bondi-diagram}}

\end{figure}

Having obtained the gravitational potential around a 'dumbbell' structure,
we now concern with the plasma accretion onto it. The Bernoulli integral
for the flow reads

\begin{equation}
\frac{u^{2}}{2}+\frac{C_{\infty}^{2}}{\gamma-1}\left[\left(\frac{\rho}{\rho_{\infty}}\right)^{\gamma-1}-1\right]+\Phi=0\label{eq:Bernoul}\end{equation}
Using the cylindrical coordinates, $u=\sqrt{u_{r}^{2}+u_{z}^{2}}\simeq u_{r}$
is the gas flow speed, $\rho$ is its density with $\rho=\rho_{\infty}$
at $r=\infty$, $\gamma$ is the gas adiabatic index, and $C_{\infty}=\sqrt{\gamma P_{\infty}/\rho_{\infty}}$
is the sound velocity at infinity. A brief comment about the overall
geometry of the flow is in order here. If we consider the dumbbell
as an isolated structure, then the flow speed at infinity $u\propto1/r^{2}$.
If, on the other hand, the dumbbell is part of an extended elongated
structure, or if the sonic point (Bondi radius) at $r=R_{B}$ is located
not far away from the filament, $R_{B}\sim l$, then the relation
$u\propto1/r$ better represents the flow at large distances. To encompass
the both options we set the accretion flux

\begin{equation}
\rho u_{r}r^{d-1}=J=\dot{M}/2\left(d-1\right)\pi\label{eq:accrrate}\end{equation}
where $\dot{M}$ is the spherical accretion rate ($d=3$) or the accretion
rate per unit length of the filament ($d=2$). Using the results of
the preceding subsection, the Bernoulli integral given by eq.(\ref{eq:Bernoul})
can be re-written as follows

\begin{equation}
\frac{1}{2}\mathcal{M}^{4/\left(\gamma+1\right)}+\frac{1}{\gamma-1}\mathcal{M}^{-2\left(\gamma-1\right)/\left(\gamma+1\right)}=\lambda\left(\frac{r}{l}\right)^{\mu}\left(1+\zeta\sinh^{-1}\frac{l}{r}\right)\label{eq:Bernoul2}\end{equation}
where we have used the following notations: $\mathcal{M}=\left(u_{r}/C_{\infty}\right)\left(\rho_{\infty}/\rho\right)^{\left(\gamma-1\right)/2}$
is the local Mach number,

\begin{equation}
\lambda=\frac{1}{\gamma-1}\left[\frac{C_{\infty}\rho_{\infty}}{u_{{\rm B}}\rho_{{\rm B}}}\left(\frac{l}{R_{{\rm B}}}\right)^{d-1}\right]^{2\left(\gamma-1\right)/\left(\gamma+1\right)},\label{eq:Bern2Nots}\end{equation}

\begin{equation}
\mu=2\left(d-1\right)\frac{\gamma-1}{\gamma+1},\;\;\;\zeta=\left(\gamma-1\right)\frac{GM_{{\rm f}}}{lC_{\infty}^{2}}=\theta_{*}\frac{\gamma-1}{\gamma_{{\rm DM}}-1}\frac{C_{{\rm DM}}^{2}}{C_{\infty}^{2}}.\label{eq:zetamu}\end{equation}
and where $u_{{\rm B}}=u\left(R_{{\rm B}}\right)$, $\rho_{{\rm B}}=\rho\left(R_{{\rm B}}\right)$.
The contour plot of eq.(\ref{eq:Bernoul2}) is shown in Fig.\ref{fig:Bondi-diagram}.
As usual for the Bondi accretion, the flow passes through the sonic
point $R_{{\rm B}}$ after which a shock may form. Evidently, our
simplified model does not allow us to find the shock position accurately.
Appealing to the $\Lambda{\rm CDM}$ simulations \citet{Miniati01,RyuKangTW03,Pfrommer06,Brunetti09}
we may expect that the most probable shock strength in terms of its
upstream Mach number is $\mathcal{M}_{1}\sim2-3$, as the strong shocks
are located well outside of the region of our interest. The flow Mach
number behind the shock can be determined using the standard formula

\[
\mathcal{M}_{2}^{2}=\frac{\gamma-1+2/\mathcal{M}_{1}^{2}}{2\gamma-\left(\gamma-1\right)/\mathcal{M}_{1}^{2}}\]
which allows us to determine the flow density and speed also for $r<R_{{\rm B}}$
using the Bernoulli equation (\ref{eq:Bernoul2}). In particular,
for sufficiently small $r$, where $\mathcal{M}\ll1$, the following
simple expression for the plasma density may be used (see eq.{[}\ref{eq:Bernoul}{]})

\begin{equation}
\frac{\rho}{\rho_{\infty}}\simeq\left[1+\zeta\sinh^{-1}\frac{l}{r}\right]^{1/\left(\gamma-1\right)}\label{eq:Rhoasy}\end{equation}
This representation of $\rho$ can be used outside of the filament,
$r>R_{{\rm f}}$ while within the DM filament the situation is somewhat
more complicated. Of course, the 'vacuum' gravitational potential
$\Phi\propto\sinh^{-1}\left(l/r\right)$ should be replaced by $\Phi\propto\theta\left(r\right)$,
according to eq.(\ref{eq:gravpot}). At the same time, when the flow
approaches the filament axis, it also turns into $z$-direction so
that $u_{z}$ component cannot be neglected compared to $u_{r}$ in
eq.(\ref{eq:Bernoul}). Clearly, our simplified treatment needs to
be modified by addressing the flow in $r$ and $z$ variables in this
case. However, we limit our consideration to the case when the particle
gyroradius $r_{g}>R_{{\rm t}}$, where $R_{{\rm t}}$ is the radius
where $u_{z}\sim u_{r}$. Aiming at our treatment of the acceleration
mechanism in Sec.\ref{sec:Acceleration}, we may expand the density
profile in eq.(\ref{eq:Rhoasy}) as

\begin{equation}
\rho/\rho_{\infty}\simeq\left(1+R_{{\rm B}}/r\right)^{1/\left(\gamma-1\right)},\label{eq:rhotorhoing}\end{equation}
which is valid for not too small $r\gtrsim l$, and where $R_{{\rm B}}$
is specified as follows

\begin{equation}
R_{{\rm B}}=\left(\gamma-1\right)GM_{{\rm f}}/C_{\infty}^{2}\gg R_{{\rm f}}>R_{{\rm t}}.\label{eq:BondiRad}\end{equation}
The simplification of the density profile using eq.(\ref{eq:rhotorhoing})
is roughly in agreement with eq.(\ref{eq:Rhoasy}) even for small
$r\ll l$, in the case of a softer but not unreasonable equation of
state, $\gamma=7/5$. For example, if $r/l$ varies between 4 and
0.04, $\rho/\rho_{\infty}$ grows from $\sim10$ to $10^{4}$ for
$\zeta=10$. Since the magnetic field component that is parallel to
the filament is compressed as the density, the above three order of
magnitude density compression over the two order of magnitude distance
variation is consistent with the field compression between the nano-gauss
intergalactic and micro-gauss intracluster field.

The parameter $\zeta$ in eq.(\ref{eq:zetamu}), which can also be
represented as $\zeta=R_{{\rm B}}/l$, regulates the total density/magnetic
field compression between the IGM and the acceleration termination
zone at $r\sim R_{{\rm t}}.$ There are some observational constraints
on $C_{{\rm DM}}$ in eq.(\ref{eq:zetamu}) being $C_{{\rm DM}}\lesssim C_{{\rm IC}}$
\citep{DMtemperature04}, where $C_{{\rm IC}}$ is the intracluster
gas thermal velocity. Thus, the value of $\zeta$ may depend on how
cool the IGM gas is compared to the IC gas. This, in turn, would depend
on whether strong external \citep{Miniati00,RyuKangTW03} accretion
shocks are present outside of $R_{{\rm B}}$ and heat the gas and
it might be that, at least in some cases, $\zeta\sim1$ or even smaller
if cooling occurs in the shocked flow \citep{FabianCoolFlows94}.
On the other hand, the expression for $R_{{\rm B}}$, if the accretion
flow originates at a distance $\gtrsim l$, should be modified by
replacing $M_{{\rm f}}$ in eq.(\ref{eq:BondiRad}) by the total mass
of the 'dumbbell' as an accreting entity. This can considerably increase
$R_{{\rm B}}$ and $\zeta$. Equally important may be strong modification
of the external shocks by accelerated CRs, e.g., \citep{mdv00}, so
that the 'subshock' Mach number would decrease substantially and so
the shock heating rate would drop as well, e.g. \citep{m97a,DruryHeating09}.

\subsection{Magnetic field around filaments and nodes\label{sub:Magnetic-field-around}}

We assume that the magnetic field is passively transported from the
filament surroundings with the accretion flow. For simplicity, we
consider the field profile created by this process around each of
the two nodes and around the filament separately. This will provide
some insight into how the field around the dumbbell structure may
be organized. Given the initial field distribution $\mathbf{B}\left(\mathbf{r},0\right)$
and the velocity field $\mathbf{u}\left(\mathbf{r},t\right)$, the
magnetic field $\mathbf{B}\left(\mathbf{r},t\right)$ can be obtained
in terms of Lagrangian variables \citet{Moffatt78}.

\subsubsection{Magnetic Field around Nodes}

Starting with the node accretion and using the spherical coordinates
$r,$$\vartheta,$$\phi$ centered in one of the nodes, we assume
that the magnetic field and the flow velocity can be represented as
$\mathbf{B}\left(r\right)=\left(B_{r,}B_{\vartheta},0\right)$ and
$\mathbf{u}\left(r\right)=\left(u_{r},0,0\right)$, respectively.
Note that the polar axis here may or may not coincide with the direction
of the filament. Using the field induction equation $\mathbf{B}_{t}=\nabla\times\mathbf{u}\times\mathbf{B}$,
we obtain the following two equations for $B_{r}$ and $B_{\vartheta}$
components 

\[
\frac{d}{dt}r^{2}B_{r}=0;\;\;\;\frac{d}{dt}ru_{r}B_{\vartheta}=0\]
with

\begin{equation}
\frac{d}{dt}\equiv\frac{\partial}{\partial t}+u_{r}\frac{\partial}{\partial r}.\label{eq:char}\end{equation}
We assume the magnetic field being initially constant everywhere and
directed along the polar axis. Therefore, at $t=0$ we have $B_{r}\left(0\right)=B_{0}\cos\vartheta$
and $B_{\vartheta}\left(0\right)=-B_{0}\sin\vartheta$. Introducing
a new variable 

\begin{equation}
\tau=-\intop_{0}^{r}\frac{dr}{u_{r}\left(r\right)},\label{eq:tau}\end{equation}
we can write the solution for $\mathbf{B}$ as follows

\begin{equation}
B_{\vartheta}\left(t,r,\vartheta\right)=-B_{0}\frac{\mathcal{F}\left[\tau\left(r\right)+t\right]}{ru_{r}\left(r\right)}\sin\vartheta\label{eq:Bth}\end{equation}

\begin{equation}
B_{r}\left(r,t,\vartheta\right)=B_{0}\frac{\mathcal{G}\left[\tau\left(r\right)+t\right]}{r^{2}}\cos\vartheta\label{eq:Br}\end{equation}
where $\mathcal{F}$ and $\mathcal{G}$ are defined by the following
relations: $\mathcal{F}\left[\tau\left(r\right)\right]=ru_{r}\left(r\right)$,
$\mathcal{G}\left[\tau\left(r\right)\right]=r^{2}$. Time asymptotically
($t\to\infty$, $r<\infty$) the field becomes purely radial almost
everywhere (except for the equatorial plane $\vartheta=\pi/2$). The
reason for such behavior is that $\mathcal{F\left(\tau\right)}$ reaches
its maximum at $\tau\sim$$\tau_{{\rm B}}\equiv\tau\left(R_{{\rm B}}\right)$
and then decays, since $u_{r}\sim1/r^{2}$ as $r\to\infty$, whereas
$\mathcal{G}\left(\tau\right)$ is unbounded. From the physical point
of view, the effective radius from which the material is accreted
should be limited by $r\lesssim R_{{\rm B}}$ or, equivalently, the
argument in expressions given by eqs.(\ref{eq:Bth}-\ref{eq:Br})
should be limited by $\tau\left(r\right)+t\lesssim\tau_{{\rm B}}$.
In addition, the magnetic field compression outside of the sonic radius
$r=R_{{\rm B}}$ is not significant. Further details about the magnetic
field geometry depend on whether or not shocks are formed in the flow.
It should be noted that spherically accreting monoatomic gas cannot
become supersonic since the sonic point $R_{{\rm B}}\to0$ for $\gamma=5/3$
\citet{Bondi52,McCrea56} (this may be seen from the Bernoulli equation).
However, there are possibilities for the gas to be shocked. First,
a diatomic index $\gamma=7/5$ may be a better choice than the $\gamma=5/3$
index. Second, there may be a softer equation of state, $\gamma<5/3$,
if there is a significant component of accelerated cosmic rays \citet{Ensslin97,KangJones05}
so that the relativistic gas index $\gamma=4/3$ may be a better approximation.
If the flow does become supersonic, we can estimate from eq.(\ref{eq:Bernoul})
$u_{r}\propto\Phi^{1/2}\sim r^{-1/2}$ (free fall accretion regime).
If, on the other hand, the flow remains ($\gamma=5/3)$ or becomes
$(\gamma<5/3$) subsonic after being shocked, we can write $u_{r}\propto r^{\left(3-2\gamma\right)/\left(\gamma-1\right)}$
and $B_{\vartheta}\left(r,\vartheta\right)=B_{\vartheta}\left(r_{{\rm sh}},\vartheta\right)\left(r_{{\rm sh}}/r\right)^{\left(2-\gamma\right)/\left(\gamma-1\right)}$,
where $r_{{\rm sh}}$ is the shock stand-off radius and $B_{\vartheta}\left(r_{{\rm sh}},\vartheta\right)$
is the magnetic field behind the shock. The latter can be obtained
using the standard shock formulae, given the field ahead of the shock.
The radial component of the magnetic field is simply $B_{r}\left(r,\vartheta\right)=B_{r}\left(r_{{\rm sh}},\vartheta\right)\left(r_{{\rm sh}}/r\right)^{2}$.
We may see that in the both cases the magnetic field becomes essentially
radial due to the conservation of the flux $r^{2}B_{r}$ in converging
flow.

\subsubsection{Magnetic field around filament}

Turning to the filament part of the flow, we use the cylindrical coordinate
with an axis $z$ along the filament and assume that the initial magnetic
field has all three components $\mathbf{B}=\left(B_{r},B_{\vartheta},B_{z}\right)$.
From the induction equation with a cylindrical flow in radial direction,
$\mathbf{u}=\left(u_{r},0,0\right)$ we obtain

\[
\frac{d}{dt}rB_{r}=0;\;\;\;\frac{d}{dt}u_{r}B_{\vartheta}=0;\;\;\frac{d}{dt}ru_{r}B_{z}=0\]
where, again, $d/dt=\partial/\partial t+u_{r}\partial/\partial r$.
As in the spherical case, the solutions of these equations can be
written down in terms of the characteristics given by equations formally
identical to eqs.(\ref{eq:char}-\ref{eq:tau}). For our purposes,
it is sufficient to realize that $B_{\vartheta}/B_{z}\propto r$ and
$B_{r}/B_{z}\propto u_{r}$. As far as the $B_{\vartheta}$ component
is concerned, the alignment between the magnetic field and the filament
clearly improves towards the axis. Turning to $B_{r}/B_{z}$ ratio
we note, that it grows as the flow approaches the filament upstream
of a shock, should the latter occur in the flow. At the same time,
this quantity should decay in the shocked flow, if the density (and
thus the $B_{z}$ component) grows fast enough with decreasing $r$.
As we have argued in the preceding subsection, an $\propto r^{-3/2}$
scaling of the density is a reasonable approximation. Therefore, we
deduce that $B_{r}/B_{z}\propto r^{1/2}$, which is a substantial
reorientation of the ambient magnetic field in the filament direction
during its convection into the filament.

We thus conclude this section with i.) magnetic field is compressed
near the nodes stronger than near the filament, thus creating magnetic
mirrors for particles accelerated around the filament, and ii.) the
field is well aligned along the filament, if it is reasonably aligned
with it outside of the accretion region. These results are essential
for the particle acceleration mechanism which we describe in Sec.\ref{sec:Acceleration}.

\section{Adiabatic invariant near separatrix\label{sec:Adiabatic-invariant-near}}

In order to calculate the time dependence of particle momentum at
the final stage of acceleration, we transform the expression for the
adiabatic invariant, given by eq.(\ref{eq:adinv}), to the following
form 

\begin{equation}
I=2\mathcal{P}\oint\sqrt{\lambda^{2}-\left(x-1\right)^{2}/x^{4}}xdx.\label{eq:adinv2}\end{equation}
Here we have used the magnetic field from eq.(\ref{eq:Bofr}) (with
$\nu=3/2$) along with the variable $x=2\sqrt{r}/\mathcal{P}$. We
also have introduced a new variable $\lambda=p\mathcal{P}/4$ that
takes $\lambda=1/4$ value at the separatrix. The last integral may
be done in terms of elliptic integrals but, as this expression involves
the elliptic integral of the third kind, it is not practical to do
so. Since we are primarily interested in the particle motion near
the separatrix, it is easier to describe this motion using eq.(\ref{eq:adinv2})
directly. Noting that $I\left(\lambda\right)$ has a singular contribution
from its upper limit (separatrix) $x\to2$ as $\lambda\to1/4$, we
calculate $\partial I/\partial\lambda^{2}$ instead of $I$, to make
the contribution from this point well expressed. The integral $I\left(\lambda\right)$
will be then easily restored, as we know its value at $\lambda=1/4$,
eq.(\ref{eq:Is1}). Keeping $\mathcal{P}=const$, eq.(\ref{eq:adinv2})
may be manipulated into 

\begin{equation}
\frac{\partial I}{\partial\lambda^{2}}=2\frac{\mathcal{P}}{\lambda^{3}}\intop_{a}^{1-\sqrt{1/2-a^{2}}}\frac{\left(1/2-z\right)^{3}}{\sqrt{z^{2}-a^{2}}\sqrt{z^{2}-2z+a^{2}+1/2}}dz,\label{eq:dIdlam}\end{equation}
where we have introduced another variable

\[
a^{2}=\frac{1}{4}-\lambda=\frac{1-p\mathcal{P}}{4},\]
so that the singular turning point now appears at the lower limit
of the integral as $a\to0$. Near the separatrix, i.e. for $a\ll1$,
the main contribution to the last integral can be evaluated as follows

\[
\frac{\partial I}{\partial\lambda^{2}}=\frac{\sqrt{2}}{4}\frac{\mathcal{P}}{\lambda^{3}}\cosh^{-1}\frac{1}{a}+\mathcal{O}\left(1\right)\]
Taking eq.(\ref{eq:Is1}) into account, we obtain the following expression
for the adiabatic invariant which is strictly valid for the motion
near the separatrix 

\begin{equation}
I\left(p,\mathcal{P}\right)=\mathcal{P}\left[2\left(4\ln\frac{1}{\sqrt{2}-1}-\pi\right)+\sqrt{2}\left(1-p\mathcal{P}\right)\ln\left(1-p\mathcal{P}\right)+\mathcal{O}\left(1-p\mathcal{P}\right)\right]\label{eq:Ifinal}\end{equation}
To explicitly find the time dependence of the particle momentum from
this expression, we write

\[
I\left(p,\mathcal{P}\left(t\right)\right)\approx I_{{\rm s}}=const\]
and introduce the following variable $\varepsilon$, which characterizes
the orbit proximity to the separatrix

\begin{equation}
\varepsilon\left(t\right)=\frac{I_{{\rm s}}}{\sqrt{2}}\left(p_{{\rm s}}-\frac{1}{\mathcal{P}}\right)\ll1.\label{eq:eps}\end{equation}
Now eq.(\ref{eq:Ifinal}) takes the following compact form

\begin{equation}
\left(1-p\mathcal{P}\right)\ln\left(1-p\mathcal{P}\right)+\varepsilon=0.\label{eq:eqforpoft}\end{equation}

\begin{figure}[h]
\includegraphics[bb=0bp 130bp 612bp 792bp,scale=0.7]{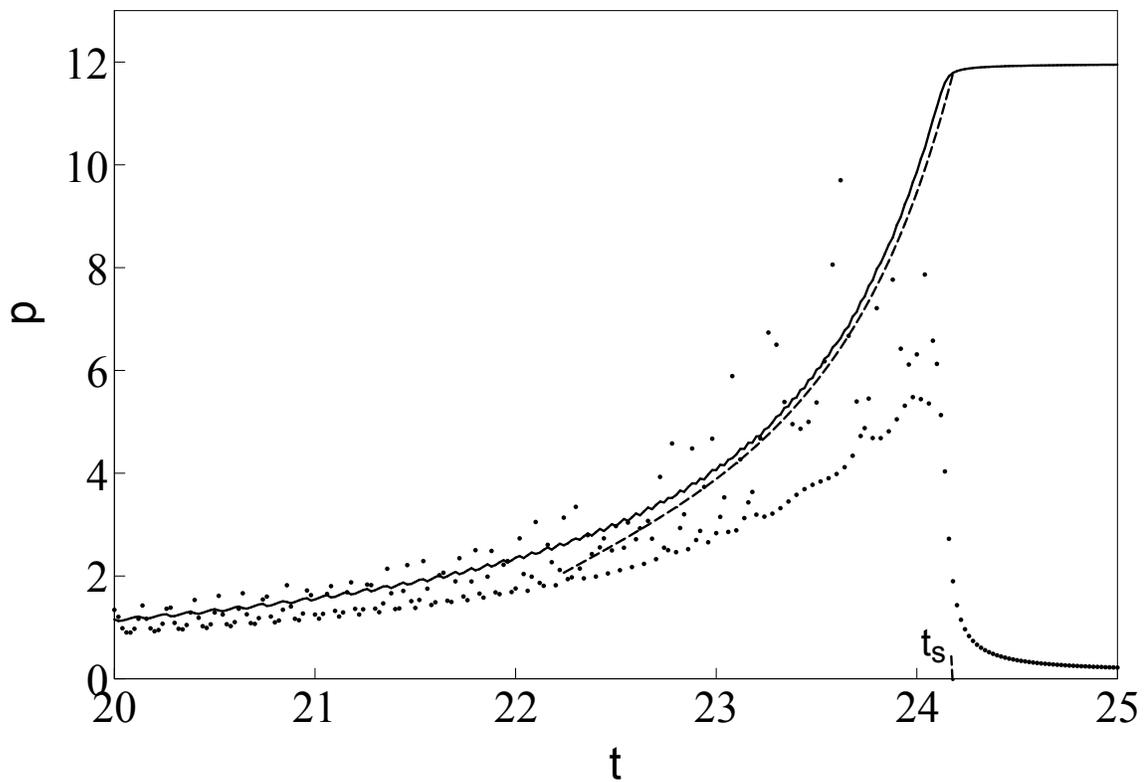}

\caption{Final phase of the acceleration process, as shown in Fig.\ref{fig:PoftWithAdiab}.
Numerical calculation of the particle trajectory is shown with the
solid line, whereas the dots represent the same (adiabatic) $p\left(t\right)$
as in Fig.\ref{fig:PoftWithAdiab}. The approximate analytic solution,
represented by eq.(\ref{eq:pOfP}), is plotted with the dashed line.\label{fig:Final-phase-of} }

\end{figure}

Recall that $\mathcal{P}$ is a known function of time: $\mathcal{P}\left(t\right)=\mathcal{P}_{0}-vt$.
Eq.(\ref{eq:eqforpoft}) can be explicitly solved for $p\left(t\right)$
using the following recursive expression

\begin{equation}
p\left(\mathcal{P}\right)=\frac{1}{\mathcal{P}}\left[1-\frac{\varepsilon}{\ln\left(\frac{1}{\varepsilon}\left(\ln\frac{1}{\epsilon}\ldots\right)\right)}\right]\label{eq:pOfP}\end{equation}
Note that this solution is valid for $1-p\mathcal{P}\leq1/e$, which
is certainly fulfilled in the case of our interest. The logarithm
recursive expression converges to the solution of eq.(\ref{eq:eqforpoft})
in this case, and it is sufficient to truncate it after 3-4 iterations.
Fig.\ref{fig:Final-phase-of} shows the final stage of acceleration,
obtained by direct numerical integration. For comparison, $p\left(t\right)$
from eq.(\ref{eq:pOfP}), truncated after four iterations, is also
shown.

As we mentioned earlier, the growth of the particle momentum with
time is explosive near the separatrix. Indeed, from the last equation
with $\varepsilon\ll1$, we obtain eq.(\ref{eq:PoftSep}). We also
note here, that in calculating the acceleration time in eq.(\ref{eq:tauacc}),
we neglected terms $\sim1/\ln\left(1/\varepsilon\right)$ compared
to unity which is consistent with the accuracy of the above calculations.

\bibliographystyle{C:/TeX/BIBS/JHEP}
\bibliography{C:/TeX/BIBS/dsa,C:/TeX/BIBS/PlasmaDSA,C:/TeX/BIBS/DSAobs,C:/TeX/BIBS/DSAshort,C:/TeX/BIBS/MALKOV,C:/TeX/BIBS/UHECR}

\pagebreak

\end{document}